\documentclass[journal,letterpaper]{IEEEtran}
\usepackage{amssymb,amsmath,amsthm}
\usepackage{amsfonts}
\usepackage{graphicx}
\usepackage{psfrag}
\usepackage{color}
\usepackage{rotating}
\usepackage[all]{xy} 

\psfrag{Crate}{$r$}
\psfrag{Qrate}{$S$}
\psfrag{R}{$R$}
\psfrag{S}{$S$}

\DeclareMathOperator{\tr}{Tr}

\DeclareMathOperator{\E}{\mathbb{E}}

\def\wh#1{{ \widehat{#1}}}

\def\NORM#1{ {\left|\hspace{-.022in}\left| #1 \right|\hspace{-.022in}\right|} }

\def\Norm#1{ {\big|\hspace{-.022in}\big| #1 \big|\hspace{-.022in}\big|} }
\def\norm#1{ {|\hspace{-.022in}|#1|\hspace{-.022in}|} }

\def\ox{\otimes}

\begin{document}
\def\pmat#1{\begin{pmatrix} #1 \end{pmatrix}}
\def\qedsymbol{\rule{7pt}{7pt}}
\def\supp{ {\rm{supp \,}}}
\def\dist{ {\rm{dist }}}
\def\dim{ {\rm{dim \,}}}
\def\oti{{\otimes}}
\def\bra#1{{\langle #1 |  }}
\def\lb{ \left[ }
\def\rb{ \right]  }
\def\tilde{\widetilde}
\def\bar{\overline}
\def\*{\star}
\def\({\left(}		\def\BL{\Bigr(}
\def\){\right)}		\def\BR{\Bigr)}
	\def\BBL{\lb}
	\def\BBR{\rb}


\def\1{{\mathbf{1} }}

\def\bb{{\bar{b} }}
\def\ab{{\bar{a} }}
\def\zb{{\bar{z} }}
\def\zbar{{\bar{z} }}
\def\inv#1{{1 \over #1}}
\def\half{{1 \over 2}}
\def\d{\partial}
\def\der#1{{\partial \over \partial #1}}
\def\dd#1#2{{\partial #1 \over \partial #2}}
\def\vev#1{\langle #1 \rangle}
\def\ket#1{ | #1 \rangle}
\def\rvac{\hbox{$\vert 0\rangle$}}
\def\lvac{\hbox{$\langle 0 \vert $}}
\def\2pi{\hbox{$2\pi i$}}
\def\e#1{{\rm e}^{^{\textstyle #1}}}
\def\grad#1{\,\nabla\!_{{#1}}\,}
\def\dsl{\raise.15ex\hbox{/}\kern-.57em\partial}
\def\Dsl{\,\raise.15ex\hbox{/}\mkern-.13.5mu D}
\def\b#1{\mathbf{#1}}
\newcommand{\proj}[1]{\ket{#1}\bra{#1}}
\def\braket#1#2{\langle #1 | #2 \rangle}
\def\1s2#1{\frac{1}{\sqrt{2^{#1}}}}
%
%
\def\th{\theta}		\def\Th{\Theta}
\def\ga{\gamma}		\def\Ga{\Gamma}
\def\be{\beta}
\def\al{\alpha}
\def\ep{\epsilon}
\def\vep{\varepsilon}
\def\la{\lambda}	\def\La{\Lambda}
\def\de{\delta}		\def\De{\Delta}
\def\om{\omega}		\def\Om{\Omega}
\def\sig{\sigma}	\def\Sig{\Sigma}
\def\vphi{\varphi}

                        \def\Up{\Upsilon}
%
%
\def\CA{{\cal A}}	\def\CB{{\cal B}}	\def\CC{{\cal C}}
\def\CD{{\cal D}}	\def\CE{{\cal E}}	\def\CF{{\cal F}}
\def\CG{{\cal G}}	\def\CH{{\cal H}}	\def\CI{{\cal J}}
\def\CJ{{\cal J}}	\def\CK{{\cal K}}	\def\CL{{\cal L}}

\def\CM{{\cal M}}	\def\CN{{\cal N}}	\def\CO{{\cal O}}
\def\CP{{\cal P}}	\def\CQ{{\cal Q}}	\def\CR{{\cal R}}
\def\CS{{\cal S}}	\def\CT{{\cal T}}	\def\CU{{\cal U}}
\def\CV{{\cal V}}	\def\CW{{\cal W}}	\def\CX{{\cal X}}
\def\CY{{\cal Y}}	\def\CZ{{\cal Z}}

\def\rvac{\hbox{$\vert 0\rangle$}}
\def\lvac{\hbox{$\langle 0 \vert $}}
\def\comm#1#2{ \BBL\ #1\ ,\ #2 \BBR }
\def\2pi{\hbox{$2\pi i$}}
\def\e#1{{\rm e}^{^{\textstyle #1}}}
\def\grad#1{\,\nabla\!_{{#1}}\,}
\def\dsl{\raise.15ex\hbox{/}\kern-.57em\partial}
\def\Dsl{\,\raise.15ex\hbox{/}\mkern-.13.5mu D}
\def\beq{\begin {equation}}
\def\eeq{\end {equation}}
\def\to{\rightarrow}
\def\h#1{\widehat{#1}}
\def\br#1{\langle #1 \rangle}

\newtheorem{lem}{Lemma}
\newtheorem{prop}{Proposition}
\newtheorem{theo}{Theorem}
\newtheorem{cor}{Corollary}

\newtheorem{rem}{Remark}
\newtheorem{dfn}{Definition}

\def\bC{\mathbb{C}}
\def\diag{\mbox{diag}}
\def\nn{\nonumber}
\def\bs#1{\boldsymbol{#1}}
\def\id{\text{\rm id}}

\def\argmax{\mbox{argmax}}
\definecolor{gray}{gray}{.9}
\def\com#1{\vspace{.1in}\fcolorbox{black}{gray}{\begin{minipage}{5.5in}#1\end{minipage}}\vspace{.1in}}

\def\deph{\Delta}

\title{Quantum Broadcast Channels}
\author{
Jon Yard$^*$ 
 \thanks{$*$  {\tt jtyard@gmail.com}, 
Computer, Computational and Statistical Sciences (CCS-3) and Center for Nonlinear Studies (CNLS), Los Alamos National Laboratory, Los Alamos, NM, USA.  Institute for Quantum Information, Caltech, Pasadena, CA, USA.  School of Computer science, McGill University, Montr\'{e}al, Canada.}
Patrick Hayden$^\ddagger$   
\thanks{$\ddagger$  {\tt patrick@cs.mcgill.ca},
School of Computer Science, McGill University, Montr\'{e}al, Canada.}
Igor Devetak$^\dagger$ 
 \thanks{$\dagger$ 
 Formerly at the Electrical Engineering Department, University of Southern California, USA.}
}

\maketitle
\thispagestyle{empty}
\begin{abstract}
We consider quantum channels with one sender and two receivers, used in several different ways for the simultaneous transmission of independent messages.  We begin by extending the technique of superposition coding to quantum channels with a classical input to give a general achievable region.  We also give outer bounds to the capacity regions for various special cases from the classical literature and prove that superposition coding is optimal for a class of channels.  We then consider extensions of superposition coding for channels with a quantum input, where some of the messages transmitted are quantum instead of classical, in the sense that the parties establish bipartite or tripartite GHZ entanglement.  We conclude by using state merging to give achievable rates for establishing bipartite entanglement between different pairs of parties with the assistance of free classical communication.
\end{abstract}


\PARstart{A}~classical discrete memoryless broadcast channel with one sender and two receivers  is modeled by a probability transition matrix $p(y,z|x)$.  Broadcast channels were introduced by Cover \cite{cover1} in 1972, and it is still not known how to compute their capacity regions in full generality.  Cover illustrated how to superimpose high-rate information on low-rate information, so that a stronger receiver obtains a refined version of what is available to a weaker receiver.
Coding theorems by Bergmans \cite{bergmans} and van der Meulen \cite{vanderM} further developed this idea, leading to the following \emph{superposition coding inner bound} to the capacity region:  Alice, the sender, can transmit a rate $R_B$ message to Bob, who sees $Y$, and a rate $R_C$ message to Charlie, who sees $Z$, while simultaneously sending a rate $R$ common message to both, if
\begin{eqnarray}
R_B &\leq& I(X;Y|T) \nn \\ 
R + R_C &\leq& I(T;Z) \label{region:csuper}\\ 
R + R_B + R_C & \leq & I(X;Y) \nn
\end{eqnarray}
for some $p(t,x)$.   Superposition coding works well when Bob receives a stronger signal than Charlie.  Making this idea precise led to the characterization of the capacity regions for several special cases.  

The first case to be solved is when Charlie's output is a \emph{degraded} \cite{cover1} version of Bob's, in which case 
the region (\ref{region:csuper}) attainable by superposition coding simplifies to 
\begin{eqnarray}
R_B &\leq& I(X;Y|T)  \nn \\
R + R_C &\leq& I(T;Z). \label{region:cdeg}
\end{eqnarray}
Cover conjectured \cite{cover1}, and Gallager proved \cite{gallager}, that this region is optimal for the degraded broadcast channel.   In Section~\ref{section:supercoding:coding}, we prove a coding theorem for quantum broadcast channels with a classical input, establishing the superposition coding inner bound (\ref{region:csuper}) for such channels.  In Section~\ref{section:supercoding:degraded}, we give an outer bound for the capacity region of degraded broadcast channels with a classical input and give conditions under which superposition coding is optimal.  Superposition coding is also optimal for classical channels in several other settings.  We recall some of these in Section~\ref{section:supercoding:outer} and illustrate how existing classical results yield outer bounds on the associated capacity regions for channels with quantum outputs.  

The remainder of the paper considers broadcast channels with a quantum input, as modeled by completely-positive trace-preserving maps.  We consider several variants that involve quantum communication.  Quantum communication is analogous to sending private information over a classical broadcast channel, where in addition to asking that Bob can decode his message with vanishing error, we ask Alice to make it essentially impossible for Charlie learn anything about Bob's message.  This cryptographic problem was first considered by Wyner \cite{wyner} and was solved by Csiszar and K\"{o}rner \cite{bcc}, who considered a setting where Alice sends a common rate $R$ message to Bob and Charlie, and a rate $R_B$ message to Bob to be kept secret from Charlie.   They showed that a randomized variant of superposition coding is optimal for this, achieving rates $(R,R_B)$ over an arbitrary broadcast channel $p(y,z|x)$ if and only if there exists $p(t,v)p(x|v)$ such that
\begin{eqnarray}
R_B &\leq & I(V;Y|T) - I(V;Z|T) \nn \\
R &\leq & \min\{I(T;Y),I(T;Z)\}. \label{region:bcc}
\end{eqnarray}
The analogy between secure classical communication and coherent quantum information was used by Devetak \cite{dev} to rigorously prove a coding theorem for quantum communication over quantum channels.  This was done by first developing secure classical codes achieving (\ref{region:bcc}) with $R=0$, then making them coherent.

In Section~\ref{section:CQforquantum}, we
give achievable rates for the analogous scenario to that of Csiszar and K\"{o}rner of sending a common classical message to Bob and Charlie, while also sending quantum information to Bob, in the sense of establishing bipartite entanglement.  In Section~\ref{section:QQforquantum}, we show that for isometric channels, the common classical message can be made coherent, enabling the generation of tripartite GHZ entanglement~\cite{GHZ} among the three participants.  In each case, we find a class of channels such that our codes are optimal, giving single-letter characterizations of the rate regions.  We conclude in Section~\ref{section:merging1} by giving achievable rates when all parties may communicate classically with each other for free in order to obtain various quantum correlations among themselves, providing applications of the state merging primitive \cite{merge} for quantum information.

\section{Preliminaries}
\subsection{Classical and quantum systems}
Throughout this paper, we use labels such as $A,B,C$ to refer to quantum systems, writing $\CH_A$ for the Hilbert space whose unit vectors correspond to the pure states of the quantum system $A$.  All Hilbert spaces will be finite dimensional, and we abbreviate $\dim\CH_A$ as $|A|$, so that $\CH_A \equiv \mathbb{C}^{|A|}$.  Given two systems $A$ and $B$, the pure states of their composite system $AB$ correspond to unit vectors in
$\CH_{AB} \equiv \CH_A\otimes\CH_B$.  When we introduce a pure state, we use a superscripted label to identify the system to which the state refers.  For 
example, $\ket{\phi}^A \in \CH_A$ and $\ket{\psi}^{AB} \in \CH_{AB}$.  The same convention will be followed when the state of a quantum system $A$ is described by a density matrix, so that $\rho^{A}\in \mathbb{C}^{|A|\times |A|}$ is a nonnegative definite Hermitian matrix with $\tr\rho^A = 1$.  For a multipartite density matrix $\rho^{ABC},$ we frequently abbreviate its
partial traces as $\rho^{AB} = \tr_C \rho^{ABC}$.  In later references to the global state, we may however omit the superscript.  We often use the abbreviation $\phi \equiv \proj{\phi}$ when referring to the rank-one density matrix corresponding to a pure state vector $\ket{\phi}$.  

For a square matrix $M$, its \emph{trace norm} $\norm{M}_1 = \tr \sqrt{MM^\dag}$ is equal to the sum of its singular values.  Given two states $\rho^A$ and $\sigma^A$, their \emph{trace distance} $\norm{\rho-\sig}_1$ is the trace norm of their difference. We use the squared version of the \emph{fidelity}, defined as $F(\rho,\sig) = \norm{\sqrt{\rho}\sqrt{\sig}}_1^2$.  When $\rho = \ket{\phi}$, the fidelity evaluates to $F(\ket{\phi},\sig)= 
\bra{\phi}\sig\ket{\phi}$.  These distances are related \cite{fuchs} via 
\begin{eqnarray}
F(\rho,\sig) &\geq& 1-\norm{\rho-\sig}_1 \label{tr2fid}\\
\norm{\rho - \sig}_1 &\leq& 2\sqrt{1-F(\rho,\sig)}. \label{fid2tr}
\end{eqnarray}  
Since the trace distance comes from a norm, it satisfies the triangle inequality
\[\norm{\rho_1 - \rho_3}_1 \leq \norm{\rho_1 - \rho_2}_1 + \norm{\rho_2 - \rho_3}_1.\]

We shall frequently make use of classical-quantum states and classical-quantum channels \cite{dcr} in this paper.  To any finite set $\CX$, we associate a Hilbert space $\CH_X$ with orthonormal basis $\{\ket{x}^X\}_{x\in \CX}$, so that for any
classical random variable $X$ which takes the value $x\in \CX$ with probability $p(x)$, we may write a density matrix
\[\rho^X = \sum_x p(x) \proj{x}^X \equiv \bigoplus_x p(x)\]
which is diagonal in that basis.
For any $\CS\subseteq \CX$, writing $P_{\CS}$ for the projector onto the subspace spanned by $\{\ket{x}^X\}_{x\in \CS}$, we then have
\[\Pr\{X\in S\} = \tr P_\CS \rho^X = \sum_{x\in \CS}p(x).\]
An ensemble of quantum states $\{\rho_x^B,p(x)\}$ can be represented in a similar way with a block diagonal \emph{classical-quantum (cq) state}
\[\rho^{XB} = \sum_{x}p(x)\proj{x}^X\otimes\rho_x^{B} \equiv \bigoplus_x p(x) \rho_x^{B}.\]
Wherever possible, we will adopt this more compact direct sum notation for describing cq states, with the understanding that the labels of the blocks correspond to states of an additional classical system.

\subsection{Quantum channels}
A \emph{classical-quantum (cq) channel} $W^{\CX\to B}$ describes a physical setup in which the sender Alice is able to remotely prepare any one of a collection of \emph{conditional density matrices} $\{\rho^B_x\}_{x\in\CX}$ in the laboratory of Bob.  By a \emph{cq broadcast channel} $W^{\CX\to BC}$ from Alice to Bob and Charlie, we mean a physical scenario in which Alice prepares any one of a collection of bipartite conditional density matrices $\{\rho_x^{BC}\}_{x\in \CX}$.

By a \emph{quantum channel} $\CN^{A'\rightarrow B}$ from $A'$ to $B$, we mean a trace-preserving linear map from density matrices on $A'$ to those on $B$ which is also \emph{completely positive}.  
Here, we parallel the state convention by treating the superscript $A'\rightarrow B$ as a definition of the domain and range of the channel, to be omitted in later references to $\CN$.  In this paper, a \emph{quantum broadcast channel} $\CN^{A'\rightarrow BC}$ refers to a quantum channel with a single input and two outputs.  We often personify the users of the channel, saying that Alice controls the input, while Bob and Charlie are located at the respective outputs.   Define the channel from Alice to Bob as $\CN^{A'\to B} \equiv \tr_C\CN^{A'\to BC}$, with a similar definition for $\CN^{A'\to C}$.  Here, the partial trace $\tr_C$ is considered as a channel from $DC$ to $D$, for any possible quantum system $D$.  One can then regard $\tr_C \CN$ as the composition of two channels.  We will say that the broadcast channel $\CN^{A\to BC}$ is \emph{degraded} whenever there exists a \emph{degrading channel} $\CN_d^{B\to C}$ from Bob to Charlie satisfying
$\CN^{A'\to C} = \CN_d^{B\to C}\circ\CN^{A'\to B}$. In other words, the following diagram must commute:
\[
\xymatrix@C=.5in{
A' \ar[r]^<>(.55){\CN^{A'\to B}} \ar[dr]_{\CN^{A'\to C}} & B \ar[d]^{\CN_d^{B\to C}}\\
& C .
}\]
We remark that
in the classical literature, such channels have been called \emph{stochastically degraded}, 
meaning that the random variables $X$, $Y$ and $Z$, 
analogous to $A$, $B$ and $C$ of the state $\rho^{ABC} = \CN^{A' \to BC}(\phi^{AA'})$,
form a Markov chain $X-Y-Z$.
However, in a quantum Markov chain \cite{hjwp} $A-B-C$ with state $\rho^{ABC}$,
there must exist a recovery map $\CM^{B\to BC}$ satisfying $\CM(\rho^{AB}) = \rho^{ABC}$.  
In our case we have the weaker condition $\CN_d^{B\to C} (\rho^{AB}) = \rho^{AC}$.
The two conditions are equivalent in the classical problem because
classical information can be copied.  

By an \emph{isometric channel}  $\CU^{A'\to B}$, we mean one given by conjugation by an isometry $U^{A \to B}$, so that $\CU(\rho) = U\rho U^\dagger$,
Given a channel $\CN^{A'\to B}$, there always exists an isometric channel $\CU^{A'\to BE}$ into an unobservable environment that \emph{extends} the channel, meaning that
$\CN^{A'\to B} = \tr_E\CU^{A'\to BE}$.  We will call such an isometry an \emph{isometric extension} of $\CN^{A'\to B}$.  While there are generally many choices for an isometric extension of a given channel, all are related via isometries on the environment $E$.  On the other hand, any channel obtained by disregarding the output $B$ of such an isometric extension will be said to be \emph{complementary} to $\CN^{A'\to B}$, which we  write  $\CN_c^{A'\to E} = \tr_B\CU^{A'\to BE}$.
In case the isometric extension $\CU^{A'\to BE}$ of $\CN^{A'\to B}$ is a degraded broadcast channel, in the sense that there is a degrading channel $\CN_d^{B\to E}$ for which
$\CN_c^{A'\to E} = \CN_d^{B\to E}\circ\CN^{A'\to B}$, we will say that the channel $\CN^{A'\to B}$ is \emph{degradable} \cite{devshor}.  Concrete examples of degradable channels include erasure channels \cite{bds}, qubit flip channels, photon number splitting channels \cite{masahito},
cloning channels and Unruh channels~\cite{BDHM,BHTW}.

A particular class of degradable channels which are relevant to this paper are the \emph{generalized dephasing channels} \cite{devshor, qmac}.  These are channels $\CN^{A'\to B}$ with $|A| = |B|$ which act noiselessly on some common orthonormal basis $\{\ket{x}^A,\ket{x}^B\}$.  Such channels have an isometric extension
\[\CU^{A'\to BE} = \sum_x \ket{\phi_x}^E\ket{x}^B\bra{x}^{A'}\]
for some (not necessarily orthogonal) normalized vectors $\{\ket{\phi_x}^E\}$, and a complementary channel acting as
\[\CN_c(\rho) = \sum_x \bra{x}\rho\ket{x} \phi_x^E.\]
Writing
\begin{equation}\Delta^{A'\to B} \colon \rho \mapsto \sum_x\proj{x}\rho\proj{x} \label{completelydephasing}\end{equation}
 for the \emph{completely dephasing channel}, which sets to zero all off-diagonal matrix elements, any generalized dephasing channel $\CN^{A'\to B}$ satisfies
\begin{eqnarray}
\CN_c\circ\Delta &=& \CN_c \label{gd1} \\
H\big(\Delta(\rho)\big) &\geq& H\big(\CN(\rho)\big) \label{gd2}.
\end{eqnarray}

For the decoding of classical information, we use (somewhat interchangably) the notions of POVM's and \emph{quantum instruments} $\bs{\CD}^{A\to BX}$.  The latter is a quantum channel whose target is a cq system.  Such a map can be specified in terms of a collection of (generally) trace-reducing maps $\{\CD_x^{A\to B}\}$ for which $\sum_x\CD_x$ is trace-preserving. The instrument then acts as
$\bs{\CD}(\rho^A) = \bigoplus_x \CD_x(\rho^A).$
Given a POVM $\{\Lambda_x\}$ on $A$, its \emph{associated measurement instrument} $\bs{\CD}^{A\to X}$ has components acting as
$\CD_x(\rho^A) = \tr\Lambda_x\rho$.

\subsection{Entropy and information quantities}
Let $\rho^{ABC}$ be any tripartite density matrix.  We write 
$H(A)_\rho = H(\rho^A)\equiv -\tr(\rho^A\log\rho^A)$ for the \emph{von Neumann entropy} of the reduced density matrix $\rho^A$, omitting the subscripted state when it is apparent.
As is common with much of quantum Shannon theory, certain linear combinations of entropies of various subsystems of the joint state $\rho^{ABC}$ arise naturally in the characterizations of the various rate regions we will introduce.  We review the essential ones here, beginning with the \emph{conditional entropy}
\[H(A|B) = H(AB) - H(B).\]
This quantity is defined in direct analogy to its counterpart in classical information theory, in which context it is always positive and can be regarded as an average entropy of conditional probability distributions.  On the other hand, $H(A|B) = -1$ when evaluated on an EPR state $\frac{1}{\sqrt{2}}\ket{00}^{AB}+ \frac{1}{\sqrt{2}}\ket{11}^{AB}$.  The negative of conditional entropy has been defined as the \emph{coherent information}
\[I(A\,\rangle B) = -H(A|B)\]
\emph{from} $A$ \emph{to} $B$, due to its utility in characterizing the capacity of a quantum channel for transmitting coherent quantum information \cite{lloyd,shor,dev} as a certain optimization problem that always yields a nonnegative rate.
Following \cite{sn}, we will sometimes use the notation
\[I_c(\rho^{A'}, \CN^{A' \to B}) \equiv I(A \, \rangle B)_{\CN(\varphi)},\] where
$\ket{\varphi}^{AA'}$ is any purification of $\rho^{A'}$.
An operational interpretation of both positive and negative conditional entropies
was found in \cite{merge}, where the primitive of \emph{state merging} was introduced, yielding a quantum counterpart to the classical Slepian-Wolf theorem for distributed data compression.  We will use this merging primitive in Section~\ref{section:merging1} to transmit quantum information over a broadcast channel.

Mutual information and conditional mutual information are respectively defined as
\[I(A;B) = H(A) - H(A|B) = H(A) + H(B) - H(AB)\]
and
\begin{eqnarray}
I(A;B|C) &=& H(A|C) - H(A|BC) \label{condmutual}\\
&=& I(A;BC) - I(A;C) \nn \\
&=& I(AC;B) - I(B;C). \nn
\end{eqnarray}
By the \emph{strong subadditivity} \cite{ssuborig} of quantum entropy, it follows that mutual information and conditional mutual information are nonnegative.  There are many equivalent formulations of strong subadditivity which we will now recall.  By simple algebra,  $I(A;B|C)\geq 0$ is seen to be equivalent to the inequality  $H(A|BC) \leq H(A|B)$ which is interpreted as saying that \emph{conditioning reduces entropy} , and thus increases coherent information $I(A\,\rangle BC) \geq I(A\,\rangle B)$.  These can easily be used to derive either form of the \emph{data processing inequality}, which say that given any channel $\CN^{B\to C}$,
\begin{eqnarray}
I(A;B)_{\rho^{AB}} &\geq& I(A;C)_{\CN(\rho^{AB})} \label{dp-mi}\\
I(A\,\rangle B)_{\rho^{AB}} &\geq& I(A\,\rangle C)_{\CN(\rho^{AB})} \label{dp-ci}.
\end{eqnarray}
In other words, processing the output of a channel will never increase the mutual or coherent information over that channel.  We remark that the first inequality above includes the Holevo bound \cite{holevobound} as a special case, since a measurement can be considered as a quantum channel with a strictly classical output.
Note that (\ref{dp-ci}) can also be written
\[I_c(\rho^{A'},\CN^{A'\to B}) \leq I_c(\rho^{A'},\CM^{B\to C}\circ\CN^{A'\to B})\]
for every $\rho^{A'}$, $\CN^{A'\to B}$ and $\CM^{B\to C}$.
Finally, given a quadripartite system $A_1A_2B_1B_2$, the following inequality is implied by and also implies strong subadditivity:
\begin{eqnarray}
H(A_1A_2|B_1B_2) \leq H(A_1|B_1) + H(A_2|B_2).
\end{eqnarray}

\section{Superposition coding for classical-quantum channels} \label{section:supercoding}
In what follows, a sequence $x_1 x_2 \cdots x_n$, with each $x_i$ belonging to some set $\CX$
will be denoted by $x^n$.
Using many instances of a cq broadcast channel $W^{\CX\rightarrow BC}$, suppose that Alice wishes send personal messages to Bob and Charlie at rates $R_B$ and $R_C$, while simultaneously sending them a rate $R$ common message.  If $W$ has conditional density matrices $\rho_x^{BC}$, we define an $(R,R_B,R_C,n,\ep)$ code for $W$ to consist of an encoding $\{x^n(m,k,\ell)\in\CX^n\}$ where $(m,k,\ell)\in 2^{nR}\times2^{nR_B}\times 2^{nR_C}$, a POVM $\{\Lambda_{mk}\}$ on $B^n$
and a POVM $\{\Lambda'_{m\ell}\}$ on $C^n$  which satisfy
\[\tr\rho_{x^n(m,k,\ell)}(\Lambda_{mk}\otimes\Lambda'_{m\ell}) \geq 1-\ep\]
for every $(m,k,\ell)\in 2^{nR}\times 2^{nR_B}\times 2^{nR_C}$. (The notation $2^{nR}$ will
be used throughout as shorthand for a set of cardinality $\lfloor 2^{nR} \rfloor$.)
A rate triple $(R,R_B,R_C)$ is \emph{achievable} if there is a sequence of 
$(R,R_B,R_C,n,\ep_n)$ codes with $\ep_n\rightarrow 0$.  The \emph{classical capacity region} $\CC(W)$ of $W$ is defined as the closure of the collection of all such achievable rate triples.  We begin by stating a generalization of the classical superposition coding region (\ref{region:csuper}) described in the introduction.  It is proved in Section~\ref{section:supercoding:coding}:

\begin{theo} \label{theo:supercoding}
Given a cq channel $W^{\CX\to BC}$ with conditional density matrices $\{\rho_x^{BC}\}$, a triple of nonnegative rates $(R,R_B,R_C)$ is achievable if
\begin{eqnarray}
R_B &\leq& I(X;B|T) \nn \\ 
R + R_C &\leq& I(T;C) \label{region:qsuper}\\ 
R + R_B + R_C & \leq & I(X;B) \nn
\end{eqnarray}
for some $p(t,x)$ with $|\CT| \leq \min\{|\CX|,|B|^2 + |C|^2 - 1\}$, giving rise to the state 
\begin{equation}
\sig^{TXBC} = \bigoplus_{t,x} p(t,x) \rho_{x}^{BC}. 
\label{supercodingarise}
\end{equation}
\end{theo}

The \emph{regularization} of a region $R(W)$ of rates associated to a channel is defined as 
\[R^\infty(W) =  \bigcup_{n=1}^\infty \frac{1}{n} R(W^{\ox n}).\]
When an achievable region does not equal the full capacity region $C(W)$, it can still be the case that its regularization does.  However, the superposition coding region does \emph{not} generally regularize to $C(W)$.  For instance, if Bob's output is completely decorrelated from the input, while Charlie's is noiseless, then the theorem requires $R + R_B + R_C = 0$ even though $R_C > 0$ is
certainly possible. 
We are, however, able to prove the optimality of superposition coding for a class of degraded channels in the following theorem, proved in Section~\ref{section:supercoding:degraded}:
\begin{theo} \label{theo:cqchanneldegraded}
Suppose that the conditional density matrices $\{\rho_x^{BC}\}$ of a cq channel $W^{\CX\to BC}$ are such that their restrictions $\rho_x^{B}$ and $\rho_x^C$ satisfy $\rho^C_x = \CM(\rho^B_x)$ for some channel $\CM^{B\to C}$. Then $C(W)$ is contained in the set of triples of nonnegative rates $(R,R_B,R_C)$ such that 
\begin{eqnarray*}
R_B &\leq& I(X;B|T)_\sig \\
R + R_C &\leq& I(T;C)_\sig
\end{eqnarray*}
for some $\sig^{TXBC}$ of the form 
\begin{equation}
\bigoplus_x p(x) \rho_x^T \ox \rho_x^{BC}
\end{equation}
for some $p(x)$ and some collection of auxiliary density matrices $\rho_x^T$.  If the $\rho_x^B$ commute, then the same region with $\sig^{TXBC}$ of the form
(\ref{supercodingarise}) with $|\CT|\leq \min\{|\CX|,|B|^2\}$ coincides with the superposition coding theorem, giving a single-letter characterization of $C(W)$.
\end{theo}

Similar outer bounds hold for other related scenarios, but it is not clear if any coincide with with our superposition coding inner bound, except for the essentially classical case where each set $\{\rho_x^B\}$ and $\{\rho_x^C\}$ mutually commutes.  We comment on these scenarios below in Section~\ref{section:supercoding:outer}.
\subsection{Superposition coding inner bound} \label{section:supercoding:coding}
Here we prove a superposition coding theorem that establishes Theorem~\ref{theo:supercoding}.  Converse proofs for the other results in this section are the subject of the next subsection. 
The coding theorem relies on the following auxiliary results.  The first is an average error version of the HSW Coding Theorem for cq codes with codewords chosen i.i.d.\ according to a product distribution \cite{sw2,holcap}.
\vspace{.05in}
\begin{prop}[HSW Random Coding Theorem] \label{prop:hsw}
Given are a cq state
$\sigma^{XQ}
=\bigoplus_x p(x) \rho^Q_x$
and a rate $0\leq R < I(X;Q)_\sigma.$ For every $\epsilon > 0$, there
is $n$ sufficiently large so that if $2^{nR}$ codewords $\CC=\{X^n(m)\}$ are chosen i.i.d.\ according to the product distribution $p(x^n) = \prod_{i=1}^n p(x_i)$, corresponding to input preparations
$\rho_{x^n} = \bigotimes_i\rho_{x_i},$ there exists a decoding POVM $\{\Lambda_m\}$ on $Q^n$, depending on the random choice of codebook $\CC$, which correctly identifies the index $m$ with average probability of error less than $\epsilon,$ in the sense that
\begin{eqnarray}
\E_\CC 2^{-nR}\sum_{m=1}^{2^{nR}} \tr\rho_{X^n(m)} \Lambda_m
\geq 1-\epsilon.
\end{eqnarray}
\end{prop}
\vspace{.05in}
We will also require the following classical-quantum analog of Corollary~3.8 from \cite{ck}.  Its proof follows from standard arguments (see e.g.\ \cite{svw}) and is thus omitted.
\vspace{.05in}
\begin{prop} \label{prop:2hsw}
Let $\{\rho^{BC}_x\}_{x\in\CX}$ be a cq channel $W^{\CX\rightarrow BC}$, and let $p(x)$ and $\ep,\delta>0$ be given.  If \[0\leq R = \min\{I(X;B),I(X;C)\}-\delta\] and $n$ is large enough, there is a set of $2^{nR}$ HSW codewords $\{x^n(m)\}$, each of the same type $P$ satisfying $\norm{P-p}_1\leq \delta$, a measurement on $B^n$ with POVM $\{\Lambda_m\}$ and a measurement on $C^n$ with POVM $\{\Lambda'_m\}$
such that for each $m$,  
\[\tr (\Lambda_m\otimes\Lambda'_m)\rho_m \geq 1-\ep\]
where $\rho_m = \bigotimes_i \rho_{x_i(m)}$.
\end{prop}
\vspace{.05in}
We will also require the following lemmas
\begin{lem}[see e.g.\ Lemma~1 of \cite{qmac}]\label{lemma:special}
Given density matrices $\rho,\sigma$ and an operator $0\leq \Lambda \leq \mathbb{I}$, \[\tr\Lambda\sigma \geq \tr\Lambda\rho - \norm{\rho-\sigma}_1.\]
\end{lem}
\vspace{.1in}
\begin{lem}[Gentle measurement (average version)~\cite{wintermac}] \label{lem:gentle}
Let $\rho, \Lambda$ be random $d\times d$ matrices such that $\rho$ is a density matrix and $0\leq \Lambda \leq \mathbb{I}$ which satisfy $\E\tr\Lambda \rho \geq 1-\epsilon$.  Then
\[\E\Norm{\sqrt{\Lambda}\rho\sqrt{\Lambda} - \rho}_1 
\leq \sqrt{8\ep}.\]
\end{lem}
\vspace{.1in}

\noindent {\bf Proof of Theorem~\ref{theo:supercoding}} ({Superposition coding theorem}).
To show that $(R,R_B,R_C)$ is achievable, it suffices to show achievability for  $(R+R_C,R_B,0)$, because if Bob ignores part of the common message, we can consider it as being intended only for Charlie.  Further note that because 
\[I(X;B) = I(TX;B) = I(T;B) + I(X;B|T),\]
the region (\ref{region:qsuper}) with $R_C = 0$ is equivalent to 
\begin{eqnarray}
R_B &\leq& I(X;B|T) \\ 
R  &\leq& \min\{I(T;B),I(T;C)\}. \label{region:qsuper2}
\end{eqnarray}
We therefore prove achievability with $R_C = 0$, focusing on rates $(R,R_B)$ of the above form.
Let $W^{\CX\to BC}$ be a cq broadcast channel with conditional density matrices $\rho_x^{BC}$ and let $p(t,x)$ be arbitrary.  Together, these probabilities and states define the joint cq state 
\begin{eqnarray}
\sig^{TXBC} 
&=& \bigoplus_{t,x}p(t,x)\rho_x^{BC}
\equiv \bigoplus_t p(t)\sig_t^{XBC}. \nn \label{sigu}
\end{eqnarray}
The corresponding conditional distribution $p(x|t)$ defines a set of conditional density matrices 
\[\tau_t^{BC} = \sum_x p(x|t)\rho_x^{BC} = \tr_X\sig_t^{XBC}\] 
for a new cq channel $V^{\CT\to BC}$, 
representing a ``backed up" version of the original channel $W$.    
Note that these conditional density matrices can be used to rewrite 
\[\sig^{TBC} = \tr_X\sig^{TXBC}  = \bigoplus_t  p(t) \tau_t^{BC}.\]
For any $\ep,\delta>0$ and sufficiently large $n$, we will show that 
for rates $R_B$ and $R$ satisfying 
\[I(X;B|T)_\sig - (1+|\CX|)\delta \leq R_B < I(X;B|T)_\sig\] and \[0\leq R = \min\{I(T;B)_\sig,I(T;C)_\sig\}-\delta,\]
there exists an $(R,R_B,n,16\sqrt{\epsilon})$ code for $W^{\CX\to BC}$.

We will construct the required doubly-indexed set of codewords $\{x(m,k)\}_{m\in 2^{nR}, k\in 2^{nR_B}}$ as follows.  First, we select a rate $R$ code for the channel
$V^{\CT\to BC}$ which conveys the index $m\in 2^{nR}$ to Bob and Charlie.
Then, for each $t$, we pick a random HSW code of blocklength approximately $p(t)n$ for $W^{\CX\to BC}$ with codewords selected i.i.d.\ according to $p(x|t)$, such that if Bob knows $t$, he can decode at rates approaching $I(X;B)_{\sig_t}$.  Note that because of the randomness in this second coding layer,
the average state seen by Bob
on any channel output where the $t$'th code was used is equal to $\tau_t^B$.

To decode, Bob and Charlie first use their measurements from the common code, allowing them to identify $m$ well on average.  In addition to knowing the common message $m$, Bob then knows which instances of the channel were used with which random codes, so that he can apply an appropriate decoder, which depends on the randomness in the second coding layer, to learn his personalized message $k$.  Note that since
\[I(X;B|T)_\sig = \sum_t p(t) I(X;B)_{\sig_t},\]
the personal rate to Bob will be near that which is desired.  
We then infer the existence of a deterministic code with low error probability for all message pairs.

We begin by invoking Proposition~\ref{prop:2hsw} to obtain an $(R,n,\ep)$ code $\{t^n(m),\Lambda_m,\Lambda_m'\}_{m\in 2^{nR}}$ for $V^{\CT\to BC}$ with codewords of type $P$ satisfying $|P-p|_1\leq \delta$.
Recall that for each $m$, 
\begin{equation}
\tr (\Lambda_m\otimes\Lambda'_m)\tau^{B^nC^n}_m \geq 1-\ep,\label{Lambdataum}
\end{equation}
where $\tau_m = \bigotimes_i\tau_{t_i(m)}.$  

For each $t$, define the integer $n_t=nP(t)$, as well as $\ep_t = \ep P(t)$,  
$\delta_t = \delta P(t)$ and $R_t = I(X;B)_{\sig_t}- \delta_t\leq |\CX|$.
It follows from Proposition~\ref{prop:hsw} that for each $t$, there exists an $(R_t,n_t,\ep_t)$ random HSW code $\{X^{n_t}(k_t|t),\Lambda^{(t)}_k\}_{k_t\in 2^{nR_t}}$
(here, $\{X^{n_t}(k_t|t)\}$ is just a doubly indexed family of random variables)
 for the channel $W^{\CX\to B}$ to Bob that satisfies
\begin{equation}
\E 2^{-nR_t}\sum_{k_t = 1}^{ 2^{nR_t}}
\tr\rho_{k_t}^{B_t}\Lambda^{(t)}_{k_t} \geq 1-\ep_t
\label{a}
\end{equation}
 where the expectation is over the
randomness in the HSW codes. 
 Above, we have abbreviated $B_t \equiv B^{n_t}$ and taken
\[\rho_{k_t}^{B_t} = \bigotimes_{i=1}^{n_t}\rho^{B_i}_{X_i(k_t|t)}.\]
Each $X_i(k_t|t)$ is chosen independently according to $p(x|t)$, so that
 $\E\rho_{k_t} = \tau_t^{\otimes n_t}$.  Observe that by the symmetry of the random code construction, (\ref{a}) may be equivalently expressed as 
\[\E \tr\rho_1^{B_t}\Lambda^{(t)}_{1} \geq 1-\ep_t.\]
Because $R_B < I(X;B|T)$ and 
\begin{eqnarray*}
R_B &=& \sum_t \frac{n_t}{n}R_t 
= \sum_t P(t)R_t \\ 
&\geq& \sum_t p(t) R_t - \norm{P - p}_1 |\CX| \\
&\geq& I(X;B|T) - (|\CX| + 1)\delta,
\end{eqnarray*}
we may uniquely identify any message $k\in 2^{nR_B}$ for Bob with a collection of messages $\{k_t\in 2^{nR_t}\}_t.$
Recalling that all of the codewords $\{t^n(m)\}_{m\in 2^{nR}}$ are of the same type and setting $d = |\CT|$, we may assume w.l.o.g.\ that $t^n(1) = 1^{n_1}2^{n_1}\cdots d^{n_d}$, so that we may identify a collection of permutations $\{\pi(m)\colon\CT^n\to\CT^n\}$ for which $t^n(m) = \pi(m)(t^n(1))$.  By letting these permutations act on $\CX^n$ in the same way, we may define Alice's (random) encoding via
\[X^n(m,k) = \pi(m)\big(X^{n_1}(k_1|1)X^{n_2}(k_2|2)\cdots X^{n_d}(k_d|d)\big).\]
We abbreviate $\rho^{B^nC^n}_{mk} = \rho^{B^nC^n}_{X^n(mk)}$, observing that for each $k$, we have $\E\rho_{mk} = \tau_m$.

To decode, Bob first measures $\{\Lambda_m\}$ while Charlie measures $\{\Lambda_m'\}$, after which they declare their respective results to be the common message $M$.  Next, Bob will permute his $B^n$ systems according to $\pi^{-1}(m)$, obtaining a state close to $\rho^{B^n}_{1m}$.  For each $t$, he then measures each block of $n_t$ outputs with the corresponding $\{\Lambda^{(t)}_{k_t}\}$ to obtain $(k_1,\dotsc,k_t) = k$, which he declares as his personal message.  Bob's overall procedure can be summarized in terms of the POVM $\{\Lambda_{mk}\}$, defined as 
$\Lambda_{mk} = \sqrt{\Lambda_m}\Lambda_{k|m}\sqrt{\Lambda_m},$ where we take 
\[\Lambda_{k|m} = \pi(m)\Big(\bigotimes_t \Lambda^{(t)}_{k_t}\Big)\]
with $\pi(m)$ now acting to permute $B^n$ in the obvious way.
Defining
\begin{eqnarray*}
P_{mk} &=&  \tr (\Lambda_{mk} \otimes \Lambda'_m)\rho^{B^nC^n}_{mk},\\
\tilde{\rho}_{mk} &=& (\sqrt{\Lambda_m}\otimes \sqrt{\Lambda_m'})\rho^{B^nC^n}_{mk}
(\sqrt{\Lambda_m}\otimes \sqrt{\Lambda_m'}),
\end{eqnarray*}
we estimate
\begin{eqnarray*}
\E P_{mk}
&=& \E\tr\tilde{\rho}^{B^nC^n}_{mk}\Lambda_{k|m} \\
&\geq& \E\tr\rho^{B^n}_{mk} \Lambda_{k|m}
- \E\Norm{\tilde{\rho}^{B^nC^n}_{mk} - \rho_{mk}^{B^nC^n}}_1 \\
&\geq& \E\tr\rho^{B^n}_{mk}\Lambda_{k|m}  - \sqrt{8 \ep}\\
&=& \E\tr\Big(\bigotimes_t\rho^{B_t}_{1}\Big)
\Big(\bigotimes_t\Lambda^{(t)}_1\Big) - \sqrt{8\ep} \\
&=& \prod_t \E\tr\rho^{B_t}_1\Lambda^{(t)}_1 -\sqrt{8\ep} \\
&\geq& 1- \sum_t \ep_t -\sqrt{8\ep}\\
&\geq& 1- 4\sqrt{\ep}.
\end{eqnarray*}
The first inequality is by Lemma~\ref{lemma:special} and the second by
Lemma~\ref{lem:gentle}.
We may now derandomize, concluding that there is a particular value of the common randomness such that
\begin{eqnarray*}
2^{-n(R_B + R)}\sum_{m=1}^{2^{nR}}\sum_{k=1}^{2^{nR_B}} P_{mk}
&\equiv& 2^{-nR}\sum_{m=1}^{2^{nR}} \overline{P}_m \\
&\geq& 1- 4{\sqrt{\ep}}
\end{eqnarray*}
By Markov's inequality, the best half of the messages $m$ satisfy $\overline{P}_m\geq 1-8\sqrt{\ep}$.  For each of those, the best half of the corresponding $k$'s satisfy $P_{mk}\geq 1-16\sqrt{\ep}.$  By only using those $m$'s, the common rate $R$ is reduced by a negligible $\frac{1}{n}$.  For each such $m$, throwing out the worst half of the $k$'s reduces $R_B$ by the same amount.  This completes the proof. \qed
\subsection{Outer bound and converse for degraded broadcast channels}\label{section:supercoding:degraded}
\noindent {\bf Proof of Theorem~\ref{theo:cqchanneldegraded}} ({cq degraded broadcast outer bound}).
Assume $(R,R_B,R_C)$ is achievable. Then  $(0,R_B,R+R_C)$ is also achievable, and we let $\{x^n(k,m)\}$, $\{\Lambda_{k}\}$ and $\{\Lambda'_m\}$ comprise any $(0,R_B,R + R_C,n,\ep_n)$ code in the achieving sequence.  Here Bob's message is $k$, Charlie's is $m$, and we drop the common message index from our notation.  Letting
\[\Pi^{KMX^n}_{km} = \proj{m}\otimes\proj{k}\otimes\proj{x^n(k,m)},\]
we write
\begin{eqnarray*}
\om^{KMX^nB^nC^n} \!\!= 2^{-n(R + R_B+R_C)}\sum_{k=1}^{2^{nR_B}}\sum_{m=1}^{2^{n(R + R_C)}} \!\!\Pi_{km} \otimes\rho^{B^nC^n}_{km}
\end{eqnarray*}
for the state induced by selecting the messages $K$ and $M$ uniformly at random.
Let $\Omega^{KM\wh{K}\wh{M}}$ be the joint state after the decoding if Bob stores his decoded messages in $\wh{K}$ and Charlie stores his in $\wh{M}$. 
Then, for some $\ep'_n,\ep''_n,\ep'''_n\to 0$,  we have
\begin{eqnarray}
n(R + R_C) &=& H(M) \nn\\
&\leq& I(M;\wh{M})_\Om + n\ep'_n \nn\\
&\leq& I(M;C^n)_\om + n\ep'_n, \label{nR}
\end{eqnarray}
The second line is by Fano's inequality (see e.g. \cite{coverthomas})
and the third is by the Holevo bound \cite{holevobound}.  We also bound
\begin{eqnarray}
nR_B &=& H(K) \nn\\
&\leq& I(K;\wh{K})_\Om + n\ep'''_n\nn\\
&\leq& I(K;B^n)_\om  + n\ep'''_n\nn\\
&\leq& I(K;B^nM)_\om + n\ep'''_n\nn\\
&=& I(K;B^n|M)_\om + n\ep'''_n\nn\\
&\leq& I(X^n;B^n|M)_\om + n\ep'''_n. \label{nR_B}
\end{eqnarray}
The middle four lines are by 
Fano's inequality, the Holevo bound, data processing,
and the independence of $K$ and $M$. The last inequality 
uses the Markov chain $KM - X^n - B^n$.

The next step is to ``single-letterize" these bounds.  The remaining steps in this proof carry over essentially without modification from the classical proof \cite{gallager}.   To be sure, however, we go through alll the steps here in full detail.
We begin by rewriting the conditional information from (\ref{nR_B}):
\begin{eqnarray*}
I(X^n;B^n|M) \!\!\!\!&=& \!\!\!\!H(B^n|M) - H(B^n|X^nM) \\
&=&\!\!\!\! \sum_{i=1}^n\big[H(B_i|B^{i-1}M) - H(B_i|X^nB^{i-1}M)\big] \\
&=&\!\!\!\! \sum_{i=1}^n\big[H(B_i|B^{i-1}M) - H(B_i|X_iB^{i-1}M)\big] \\
&=&\!\!\!\! \sum_{i=1}^n I(X_i;B_i|MB^{i-1}) \\
&=&\!\!\!\! nI(X_S;B_S|MB^{S-1}S) \\
&=&\!\!\!\! nI(X;B|T).
\end{eqnarray*}
The third line holds because
of the Markov chain \[X^{i-1}X_{i+1}^n - X_iMB^{i-1} - B_i,\]
where we abbreviate $X_{i+1}^n = X_{i+1}\cdots X_n$ for $i < n$, setting it equal to a constant when $i=n$.
To see that this is a Markov chain, note that the left recovery map is deterministic, while the right recovery map 
prepares the appropriate state of $B_i$ given the value of $X_i$.
In the remaining steps, we let 
$S$ be uniformly distributed on $\{1,\dotsc,n\}$.  In the last step, we make the identification $T\equiv SMB^{S-1}$ and $X\equiv X_S$.  We continue by bounding the mutual information appearing in (\ref{nR}):
\begin{eqnarray}
I(M;C^n) &=& \sum_{i=1}^n I(M;C_i|C^{i-1})  \nn \\
&=& \sum_{i=1}^n \big[H(C_i|C^{i-1}) - H(C_i|MC^{i-1})\big] \nn \\
&\leq& \sum_{i=1}^n \big[H(C_i) - H(C_i|MC^{i-1})\big] \nn \\
&\leq& \sum_{i=1}^n \big[H(C_i) - H(C_i|MB^{i-1})\big] \nn \\
&=& \sum_{i=1}^n I(MB^{i-1};C_i) \nn \\
&=& nI(MB^{S-1};C_S|S) \nn \\
&\leq& n\big[I(MB^{S-1};C_S|S) + I(S;C_S)\big]\nn  \\
&=& nI(SMB^{S-1};C_S)\nn \\
&=& nI(T;C).
\label{nIMC}
\end{eqnarray}
Here, the third and fourth lines follow from the fact that conditioning reduces entropy
and data processing with respect to appropriate tensor products of the degrading map $\CM^{B\to C}$.
The last step identifies $C\equiv C_S$.
Observe that $T$ can be identified with a classical random variable only in the case where the $\{\rho^B_x \}$ are mutually commuting.  We do not know if the general outer bound with non-classical $T$ is actually achievable. \qed
\subsection{Outer bounds for other special cases}\label{section:supercoding:outer}
There are several other scenarios for which the superposition inner bound leads to single-letter characterizations of the associated capacity region over a classical broadcast channel.    A general result was given by K\"orner and Marton \cite{KM77}, who considered the special case with $R_C = 0$, where Bob is required to decode all of the information sent to Charlie.  They showed that superposition coding is optimal for this \emph{degraded message set} scenario by proving that for an arbitrary broadcast channel, the region 
\begin{eqnarray*}
R_B &\leq& I(X;Y|T) \nn \\ 
R  &\leq& I(T;Z) \\ 
R + R_B & \leq & I(X;Y) \nn
\end{eqnarray*}
obtained by specializing (\ref{region:csuper}) to $R_C = 0$ is optimal.

In \cite{EG79}, El Gamal considered two scenarios without a common message.  He showed that 
if Bob's output is \emph{less noisy} than Charlie's, meaning that $I(T;Y) \geq I(T;Z)$ for every $p(t,x)$, then the region 
\begin{eqnarray*}
R_B &\leq& I(X;Y|T)  \nn \\
R_C &\leq& I(T;Z).
\end{eqnarray*}
obtained by specializing the degraded broadcast region (\ref{region:cdeg}) to $R = 0$ is optimal.  He also showed that if Bob's output is \emph{more capable} than Charlie's, in the sense that $I(X;Y) \geq I(X;Z)$ for every $p(x)$, then the region 
\begin{eqnarray*}
R_B &\leq& I(X;Y|T) \nn \\ 
R_C  &\leq& I(T;Z) \\ 
R_B + R_C & \leq & I(X;Y) \nn
\end{eqnarray*}
obtained by specializing the superposition coding region (\ref{region:csuper}) to $R = 0$ is optimal.
Note the implications 
\[\text{degraded } \Rightarrow \text{ less noisy } \Rightarrow \text{ more capable.}\]

By using similar methods as in Section~\ref{section:supercoding:degraded}, outer bounds can be given for each of these scenarios, having the exact same forms as given here.  In all cases, however, we only know how to prove the single letter converses by choosing an auxiliary system $T$ that contains the systems $B_1 \cdots B_{S-1} S C_{S+1} \cdots C_n$.  As we do not know how to achieve rates in Theorem~\ref{theo:supercoding} unless $T$ is classical, this means that we only know how to make these outer bounds tight in the essentially classical case where the sets $\{\rho_x^B\}$ and $\{\rho_x^C\}$ mutually commute.  Interesting questions include whether this outer bound can be achieved, and whether it still holds with $T$ classical.

\section{Classical-quantum region $\CC\CQ(\CN)$ for quantum channels} \label{section:CQforquantum}
We now consider a scenario in which Alice wishes to send quantum information to Bob at rate $Q$, while sending a rate $R$ common classical message to Bob and Charlie.  To this end, she prepares one of many states $\{\ket{\Upsilon_m}^{AA'^n}\}$ which are entangled between a system $A$ in her laboratory and the inputs of some large number of
 parallel identical broadcast channels.  Bob employs a quantum instrument $\boldsymbol\CD^{B^n\to \h{A}M_B}_1$, with the goal of learning the classical message, as well as holding the $\h{A}$ part of a highly entangled state.  Meanwhile, Charlie performs a measurement, modeled by the instrument $\bs{\CD}_2^{C^n\to M_C}$, to learn the common classical message.  Such components will be said to comprise an $(R,Q,n,\epsilon)$ \emph{cq entanglement generation code} for the broadcast channel $\CN^{A'\rightarrow BC}$ if, for each $m$,
\[F\Big(\ket{m}^{M_BM_C}\ket{\Phi_Q}^{A\h{A}}, (\bs{\CD}_1\otimes\bs{\CD}_2)\circ\CN^{\otimes n}\big(\Upsilon_m^{AA'^n}\big)\Big)
 \geq 1-\epsilon,\]
 where  $\ket{m}^{M_BM_C} \equiv \ket{m}^{M_B}\ket{m}^{M_C}$ and where $\ket{\Phi_{Q}}^{A\h{A}}$ is some fixed
\emph{rate $Q$ EPR state} 
\[\ket{\Phi_{Q}}^{A\h{A}} = \frac{1}{\sqrt{2^{nQ}}}\sum_{a = 1}^{2^{nQ}} \ket{a}^{A}\ket{a}^{\h{A}}.\]
A pair of nonnegative rates $(R,Q)$ is called an \emph{achievable cq rate pair for entanglement generation} if there is a sequence of $(R,Q,n,\epsilon_n)$
cq entanglement generation codes with $\ep_n\rightarrow 0$.
The \emph{cq capacity region  for entanglement generation} $\CC\CQ(\CN)$ is defined as
the closure of the set of such achievable cq rate pairs.
The following theorem gives achievable rates for this problem and characterizes $\CC\CQ(\CN)$ of any broadcast channels as a regularized union of rectangles.
\begin{theo} \label{theo:cq}
Let $\CN^{A'\rightarrow BC}$ be arbitrary.
Then $\CC\CQ(\CN)$ contains the closure of the collection of pairs of nonnegative  
cq rates $(R,Q)$ satisfying
\begin{eqnarray*}
Q &\leq& I(A\,\rangle BT)_\sigma \\
R &\leq& \min\{I(T;B)_\sigma,I(T;C)_\sigma\} 
\end{eqnarray*}
for some state
\begin{equation}
\sigma^{TABC} = \bigoplus_t p(t) \CN^{\otimes k}(\phi_t^{AA'}). \label{cqarise}
\end{equation}
arising from the action of $\CN$ on the $A'$ part of some bipartite pure state ensemble $\{p(t),\ket{\phi_t}^{AA'}\}.$
To compute the above region, it suffices for $|\CT|\leq \min\{|A'|^{2},|B|^{2} + |C|^{2} -1\}$.
Furthermore, the above region regularizes to equal the capacity region $\CC\CQ(\CN)$.
\end{theo}

Our next theorem gives a single-letter characterization of $\CC\CQ$ whenever Charlie holds part of the environment of a generalized dephasing channel from Alice to Bob.
\begin{theo} \label{theo:cqdephasing}
Let $\CN^{A'\rightarrow BC}$ have an isometric extension
\[\CU = \sum_x\ket{x}^B\ket{\psi_x}^{CE}\bra{x}^{A'}\]
so that $\CN^{A'\to B}$ is a generalized dephasing channel.
  Then $\CC\CQ(\CN)$ equals those pairs of nonnegative cq rates $(R,Q)$ satisfying
\begin{eqnarray*}
Q &\leq& H(X|T) - H(CE|T) \\
R &\leq& I(T;C)
\end{eqnarray*}
for some state 
\[\omega^{TXCE}=\bigoplus_{t,x} p(t,x)\psi_x^{CE}\]
with $|\CT|\leq |\CX|$.
\end{theo}

\begin{figure} 
\center{
\includegraphics[width=.3\textwidth]{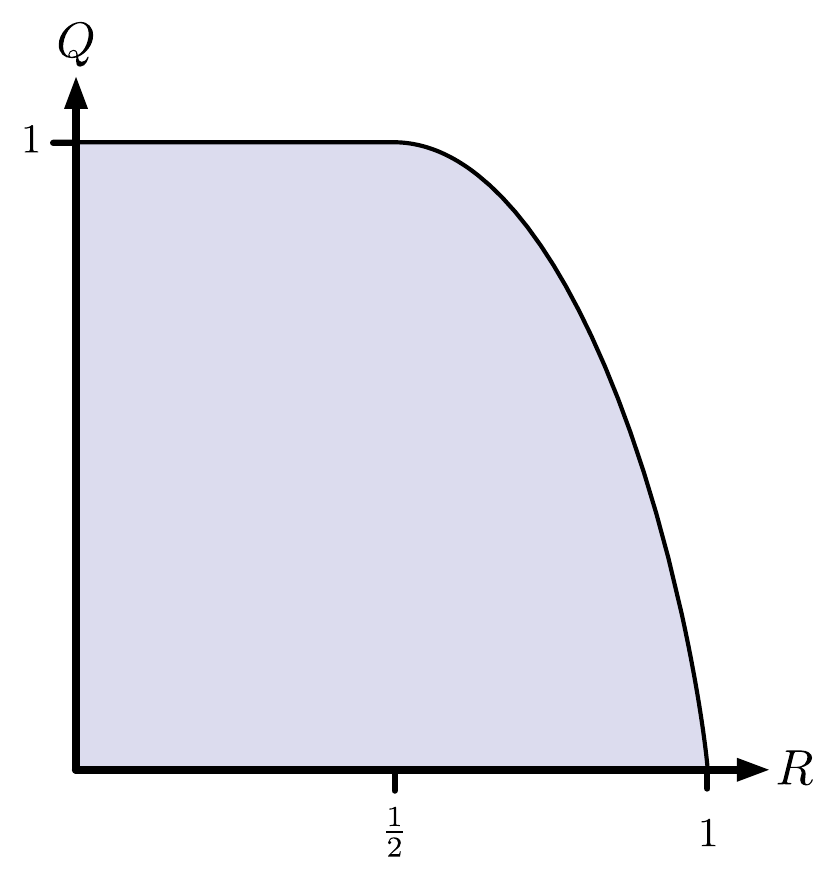}
}
\caption{Classical-quantum capacity region (\ref{eqn:pinchingregion}) for the pinching channel.} 
\label{figure:pinching}
\end{figure}

In particular, this theorem applies to any isometric extension of the following \emph{pinching channel} 
$\CP\colon \mathbb{C}^{3\times 3}\to \mathbb{C}^{3\times 3}$, which acts by setting some matrix elements to zero, while leaving the others alone, according to 
\[\CP\colon \pmat{\ast & \ast & \ast \\ \ast & \ast & \ast \\ \ast & \ast & \ast} \mapsto 
\pmat{ \ast &  \ast &    \\  \ast & \ast &  \\   &  & \ast},\] 
For the broadcast channel corresponding to any isometric extension $\CU_{\CP}^{A'\to BC}$ of $\CP$ where Charlie obtains the entire environment of $\CP$, a straightforward derivation reveals that the outer boundary of
$\CC\CQ(\CU_\CP)$ is given by  
\begin{eqnarray}
       Q &=& p \nonumber \\
       R &=& \begin{cases}
               1& \text{ if }  p \leq 1/2 \\
               H(p)& \text{ if } p \geq 1/2
             \end{cases} \label{eqn:pinchingregion}
\end{eqnarray}
where $0\leq p \leq 1$,
as is shown in Figure~\ref{figure:pinching}.

\subsection{Classical-quantum coding theorem for quantum channels}
\label{section:CQforquantum:coding}
Here we construct codes of a similar form to those used by Devetak-Shor \cite{devshor} for simultaneous transmission of classical and quantum information over single-user channels.  A basic component in our proof is the following construction of random Lloyd-Shor-Devetak (LSD) codes for entanglement generation, with average code density matrix arbitrarily close to a product state.
\vspace{.05in} 
\begin{prop}[LSD Random Coding Theorem \cite{dev}] \label{prop:lsd}
Given are a channel $\CN\colon A'\rightarrow B$, a density matrix $\rho^{A'}$, and a number $0\leq R < I_c(\rho,\CN).$  For every $\epsilon>0$, there is $n$ sufficiently large so that there is a random ensemble of $(2^{nR},n,\epsilon)$ entanglement generation codes 
$(p_\beta,\ket{\Upsilon_\beta}^{AA'^n},\CD^{B^n\to \h{A}}_\beta)$ for $\CN$
with average code density operator $\varrho^{A'^n} = \sum_\beta p_\beta\tr_{A}\Upsilon_\beta$ satisfying
$\norm{\varrho - \rho^{\otimes n}}_1\leq\epsilon.$
Moreover, each code in the ensemble is good, in the sense that
$F\big(\ket{\Phi}^{A\widehat{A}},\CD_\beta\circ
\CN^{\otimes n}(\Upsilon_\beta^{AA'^n})\big)\geq 1-\epsilon$
for each value of the randomness $\beta$.  In practice, we suppress the common randomness index from our notation, treating the encoding and decoding as random variables.
\end{prop}

\noindent {\bf Proof of Theorem~\ref{theo:cq}} ({Coding theorem}).
Let $\CN^{A'\to BC}$ be an arbitrary broadcast channel and fix an ensemble of bipartite pure states $\big\{p(t),\ket{\phi_t}^{A''A'}\big\}$.
For any $\ep,\delta>0$ and sufficiently large $n$, we will show that there exists an $(R,Q,n,12\sqrt\epsilon)$ cq entanglement generation code \[\Big\{\ket{\Upsilon_m}^{AA'^n},\bs{\CD}_1^{B^n\to M_B\h{A}},\bs{\CD}_2^{C^n\to M_C}\Big\}_{m\in 2^{nR}}\]
for $\CN^{A'\to BC}$, provided that $0\leq Q = I(A''\,\rangle BT)_\sig-\delta$ and $0\leq R = \min\{I(T;B)_\sig,I(T;C)_\sig\}-\delta$,
where
\begin{eqnarray}
\sig^{TABC}
&=& \bigoplus_{t} p(t)\CN(\phi_t^{A''A'}). \nn
\end{eqnarray}
We do this by showing that there are two POVMs: $\{\Lambda_m\}_{m\in 2^{nR}}$ on $B^n$ and $\{\Lambda'_m\}_{m\in 2^{nR}}$ on $C^n$, as well as a collection of maps $\CD_m^{B^n\to\h{A}}$, for which the trace-reducing maps  $\{\CD_m\big(\sqrt{\Lambda_m}\,(\,\cdot\,)\sqrt{\Lambda_m}\big)\}_{m\in 2^{nR}}$ are the components of $\bs{\CD}_1$, and $\bs{\CD}_2$ simply implements $\{\Lambda'_m\}_{m\in 2^{nR}}$.

For each $t$, we set $\rho_t^{A'} = \tr_{A''}\phi_t$ and $\tau_t^{BC} = \CN(\rho_t)$, defining a cq channel $V^{\CT\to BC}$ with conditional density matrices $\tau_t^{BC}$.   As in the previous coding theorem, we invoke Proposition~\ref{prop:2hsw} to obtain, for sufficiently large $n$,  an $(R,n,\ep)$ code $\{t^n(m),\Lambda_m,\Lambda_m'\}_{m\in 2^{nR}}$ for $V$ with codewords of type $P$ satisfying $\norm{P-p}_1\leq \ep$.
For each $m$, we abbreviate $\rho^{A'^n}_m \equiv \bigotimes_i\rho^{A'}_{t_i(m)}$ and recall that  
\begin{equation}
\tr (\Lambda_m\otimes\Lambda'_m)\CN^{\otimes n}(\rho_m^{A'^n}) \geq 1-\ep.
\label{lambdarhoep}
\end{equation}

As in the direct coding part of the proof of Theorem~1 we define $n_t = nP(t)$, $\ep_t = \ep P(t)$ and $\delta_t = \delta P(t)$.
We also assume that for $|\CT| = d$, the first codeword is
$t^n(1) = 1^{n_1}2^{n_2}\cdots d^{n_d}$ so that there are permutations $\pi(m)$ of $\CT^n$ satisfying $t^n(m) = \pi(m)\big(t^n(1)\big)$.

For each $t\in \CT$, we may set $Q_t = I_c(\tau_t,\CN) - \delta_t$ and conclude from Proposition~\ref{prop:lsd} that there exists a $(Q_t,n_t,\ep_t)$ random entanglement generation code $\big\{\ket{\Upsilon_t}^{A_tA'^{n_t}},\CD_t^{B^{n_t}\to\h{A}_t}\big\}$ whose average code density operator
$\varrho_t^{A'^{n_t}} = \E\tr_{A_t}\Upsilon_t$ satisfies
\begin{equation}
\Norm{\varrho_t^{A'^{n_t}} - \rho_t^{\otimes n_t}}_1\leq \ep_t. \label{avedensityept}
\end{equation}
It is also guaranteed that for each $t$, the state 
\[\xi_t^{A_tB^{n_t}C^{n_t}} \equiv \CN^{\otimes n}(\Upsilon_t)\]
 created by the $t$th random quantum code approximately contains rate $Q_t$ entanglement between Alice and Bob, in the sense that 
\begin{equation}
F\big(\ket{\Phi_{Q_t}}^{A_t\h{A}_t},\tr_{C_{n_t}}\CD_t(\xi_t)\big)
\geq 1-\ep_t. \label{Ft}
\end{equation}
Equating $A\equiv \bigotimes_t A_t$, we make the definitions
\begin{eqnarray*}
\ket{\Upsilon_1}^{AA'^n} &=& \bigotimes_t\ket{\Upsilon_t}^{A_tA'^{n_t}}\\ 
\ket{\Upsilon_m}^{AA'^n} &=& \big(1^A\otimes\pi(m)\big)\ket{\Upsilon_1}^{AA'^n},
\end{eqnarray*}
where we extend $\pi(m)$ to act by permuting the registers $A'^n$ in the obvious way.  Defining the average code density operator for the new code as $\varrho_m^{A'^n} = \E\tr_A\Upsilon_m,$ note that
we can bound
\begin{eqnarray}
\norm{\varrho_m - \rho_m}_1 &=& 
\NORM{\bigotimes_t \varrho_t^{A'^{n_t}} - \bigotimes_t \rho_t^{\otimes n_t}}_1 \nn\\
&\leq& \sum_t \norm{\varrho_t^{A'^{n_t}} - \rho_t^{\otimes n_t}}_1 \nn\\
&\leq& \sum_t \ep_t \nn\\
&=& \ep, \label{avedensityep}
\end{eqnarray}
where we have used unitary invariance of the trace norm, telescoping, and (\ref{avedensityept}), in that order. 
To send the classical message $m$, Alice prepares the state $\ket{\Upsilon_m}^{AA'^n}$.
The structure of the decoder is similar to that from the proof of Theorem~\ref{theo:supercoding}.  Bob and Charlie begin by performing their respective measurements, in order to ascertain the classical message.  Then Bob permutes his output systems accordingly and applies the quantum decoder $\CD\equiv\bigotimes_t\CD_t$. 

We will write the the joint state after Alice sends her encoding through the channel as
\[\vartheta^{AB^nC^n}_m = \CN^{\otimes n}(\Upsilon_m^{AA'^n}),\]
so that in particular, the state corresponding to the first message is $\vartheta^{AB^nC^n}_1 = \bigotimes_t \xi_t^{A_tB^{n_t}C^{n_t}}$.
Note that if the decoder $\CD^{B^n\to \h{A}}$ is applied directly to $\vartheta_1$, the resulting Alice-Bob state is nearly maximally entangled:
\begin{IEEEeqnarray}{rCl}
\IEEEeqnarraymulticol{3}{l}{
  F\big(\ket{\Phi_Q}^{A\h{A}},\tr_{C^n}\CD(\vartheta_1)\big)
}\nn\\ \qquad\qquad \qquad
&\,=\,\,\,&  \prod_t 
F\big(\ket{\Phi_{Q_t}}^{A_t\h{A}_t},\tr_{C^{n_t}}\CD_t(\xi_t)\big) \nn \\
&\,\geq\,\,\,& \prod_t (1-\ep_t) \nn \\
&\,\geq\,\,\,& 1 - \sum_t \ep_t \nn \\
&\,=\,\,\,& 1-\ep, \label{F1}
\end{IEEEeqnarray}
where the first inequality is by (\ref{Ft}).
Next, define the subnormalized density matrices 
\[\tilde{\vartheta}_m^{AB^nC^n} = 
(\sqrt{\Lambda_m}\otimes\sqrt{\Lambda'_m})\vartheta^{AB^nC^n}_m
(\sqrt{\Lambda_m}\otimes\sqrt{\Lambda'_m})\]
which depend on the shared randomness in the quantum code (as do the $\vartheta_m$), and are proportional 
to the states which result when Bob and Charlie both correctly learn the message $m$.  This happens with probability bounded as
\begin{eqnarray}
\E\tr\tilde{\vartheta}_m^{AB^nC^n} &=& 
\tr(\Lambda_m\otimes\Lambda'_m)\E\tr_A\vartheta^{AB^nC^n}_m \nn\\
&=& \tr(\Lambda_m\otimes\Lambda'_m)\CN^{\otimes n}(\varrho^{A'^n})\nn \\
&\geq& \tr(\Lambda_m\otimes\Lambda'_m)\CN^{\otimes n}(\rho^{A'^n}) \nn\\
& & - \Norm{\varrho^{A'^n} - \rho^{A'^n}}_1 \nn\\
&\geq& 1 - 2\ep.  \label{varthetas}
\end{eqnarray}
In the second to last line, we have applied Lemma~\ref{lemma:special} along with monotonicity with respect to $\CN^{\otimes n}$, 
while the last line uses the estimates (\ref{lambdarhoep}) and (\ref{avedensityep}). 
We may now write the expectation, over the shared randomness in the quantum code, of the fidelity $F_m$ between the state resulting from the protocol when the $m$th common message is sent and the target maximally entangled state as
\begin{eqnarray}
\E F_m 
&=& \E F_1 \nn\\
&\equiv& \E F\big(\ket{\Phi_Q}^{A\h{A}}, 
\tr_{C^n}\CD(\tilde{\vartheta}_1^{AB^nC^n})\big) \nn\\
&\geq& 1 - \E\Norm{\Phi_Q -  
\tr_{C^n}\CD(\tilde{\vartheta}_1)}_1 \nn \\
&\geq& 1 - \E\Norm{\Phi_Q -  
\tr_{C^n}\CD(\vartheta_1)}_1  - \E\norm{\vartheta_1 - \tilde{\vartheta}_1}_1\nn \\
&\geq& 1 - 2\sqrt{\ep} - \sqrt{8\cdot 2\ep} \nn\\
&\geq& 1-6\sqrt{\ep}. \nn
\end{eqnarray}
Here, the first line follows by the permutation symmetry of the code, while the third uses (\ref{tr2fid}).  The fourth is a consequence of the triangle inequality, together with monotonicity with respect to $\tr_{C^n}\CD$.   The estimates in the second to last line are obtained by applying (\ref{fid2tr}) to (\ref{F1}) (which holds without the expectation), as well as Lemma~\ref{lem:gentle} to (\ref{varthetas}).

At this point, it is possible to derandomize our code.  Having proved that
\[\E 2^{-nR}\sum_m F_m \geq 1-6\sqrt{\ep},\] 
we may conclude that there is a deterministic value of the shared randomness from the quantum codes yielding the same average error bound.  By throwing out the worst half of the codewords, Markov's inequality implies that we are left with a code for which 
\[F_m \geq 1-12\sqrt{\ep}\]
for each $m$, while reducing the rate by a negligible $\frac{1}{n}$. \qed
\subsection{Converse theorems}
\label{section:CQforquantum:converse}
We now complete the proof of Theorem~\ref{theo:cq} by proving a multi-letter converse to show that the regularization of the above achievable region is equal to $\CQ(\CN)$.  We require the following continuity lemma: 
\begin{lem}[Continuity] \label{lemma:continuity}
If $\Norm{\rho^{AB} - \sig^{AB}}_1\leq \delta$ for some $0\leq\delta \leq 1/e$, then
the following inequalities hold:
\begin{eqnarray*}
\big|H(A|B)_\rho - H(A|B)_\sig\big|&\leq& 2 H(\delta) + 4\delta\log|AB|\\ 
\big|I(A;B)_\rho - I(A;B)_\sig\big|&\leq& 3 H(\delta) + 6\delta\log|AB|. 
\end{eqnarray*}
\end{lem}
\vspace{.05in}
\noindent \emph{Proof.}
Fannes' \cite{Fannes73} has shown that 
\[\big|H(\rho^{AB}) - H(\sig^{AB})\big| \leq H(\delta) + 2\delta\log|AB|.\]  By monotonicity, the trace distances between partial traces of $\rho^{AB}$ are no greater than $\delta$, so after expanding the conditional entropies and mutual informations, the triangle inequality gives the result.  
\qed

\vspace{.1in}
\noindent {\bf Proof of Theorem~\ref{theo:cq}} ({Multi-letter converse}).
Assume that $(R,Q)$ is achievable and let $\{\ket{\Upsilon_m}^{AA'^n}\}_{m\in 2^{nR}}$,
$\bs{\CD}_1^{B^n \rightarrow \h{A}M_B}$ and
$\bs{\CD}_2^{C^n \rightarrow M_C}$ be a
$(R,Q,n,\ep_n)$ cq entanglement generation code from any achieving sequence.
Defining the state
\begin{equation}
\omega^{MAB^nC^n} = 2^{-nR}\bigoplus_{m\in 2^{nR}} \CN^{\otimes n}(\Upsilon_m^{AA'^n}) \label{cqconverseomega}
\end{equation}
and setting $\Omega^{MM_BM_CA\h{A}} = (\bs{\CD}_1\otimes \bs{\CD}_2)(\omega),$
we may upper bound the quantum rate $Q$ via
\begin{eqnarray}
I(A\,\rangle B^nM)_\omega &\geq& I(A\,\rangle \h{A})_\Omega \label{cqconverseIc} \\
&\geq& I(A\,\rangle \h{A})_{\Phi_{Q_B}} - n\ep'_n \nn \\
&=& nQ - n\ep_n'. \nn
\end{eqnarray}
The first step is by data processing with respect to $\tr_M\CD_1$, while the second is by the Continuity Lemma \ref{lemma:continuity}, for some $\ep'_n \to 0$.
The classical rate $R$ may also be bounded as
\begin{eqnarray}
nR &=& H(M)_\Omega \nn \\
&\leq& I(M;M_C)_\Omega \nn + n\ep_n'' \\
&\leq& I(M;C^m)_\omega  + n\ep_n'', \label{cqconverseI}
\end{eqnarray}
where $\ep''_n\to 0$, and we have used Fano's inequality  and the Holevo bound.
Another consequence of the Holevo bound is that $I(M;M_B)_\Om \leq I(M;B^n)_\om$,
yielding $nR \leq \min\{I(M;B^n)_\om, I(M;C^n)_\om\}$.
We have thus shown that for any $\delta > 0$,
the rate pair $(R-\delta,Q-\delta)$ is contained in $\CQ(\CN)$.  As $\CQ(\CN)$ is closed by definition, this completes the proof. \qed

\vspace{.1in} 

\noindent {\bf Proof of Theorem~\ref{theo:cqdephasing}} (Generalized dephasing converse). Under the assumption that $\CN^{A'\to B}$ is a generalized dephasing channel, we will further upper bound the information quantities (\ref{cqconverseIc}) and (\ref{cqconverseI}) appearing in the above multi-letter converse by appropriate single-letter quantities.  We begin working with the state $\om^{MAB^nC^n}$ from (\ref{cqconverseomega}) which is
induced by an $(R,Q,n,\ep_n)$ cq entanglement generation code from an achieving sequence.
Recalling from (\ref{completelydephasing}) that the completely dephasing channel $\Delta$
sets to zero all off-diagonal matrix elements in the dephasing basis $\{\ket{x}\}$, set $\varrho^{A'^n}_m = \deph^{\!\otimes n}(\tr_A\Upsilon_m)$, observing that we may write
\[\varrho_m^{A'^n} = \bigoplus_{x^n} p(x^n|m)\]
for some conditional probabilities $p(x^n|m)$.  Let us now
define the state
\begin{eqnarray*}
\omega'^{MB^nC^nE^n}
&=& 2^{-nR}\bigoplus_{m\in 2^{nR}}\CU^{\otimes n}(\varrho_m)\\
&=& 2^{-nR}\bigoplus_{m\in {2^{nR}}} \sum_{x^n}p(x^n|m) \psi_{x^n}^{C^nE^n}
\end{eqnarray*}
where we abbreviate $\psi_{x^n}^{C^nE^n} \equiv \bigotimes_i \psi_{x_i}^{C_iE_i}$.
Abbreviating $\CN^{A' \rightarrow B}$ to $\CN_{B}$ and $\CN^{A' \rightarrow C}$ to $\CN_{C}$,
the left hand side of (\ref{cqconverseIc}) can be written
\begin{equation}
I(A\,\rangle B^nM)_\omega
= 2^{-nR}\sum_{m\in 2^{nR}} I_c(\tr_A\Upsilon_m,\CN_B^{\otimes n}). \nn
\end{equation}
By (\ref{gd1}) and (\ref{gd2}), each summand can be bounded above as
\begin{eqnarray*}
I_c(\tr_A\Upsilon_m,\CN_B^{\otimes n})
&\leq& H\big(\CN_B^{\otimes n}(\varrho_m)\big)
- H\big((\CN_B)_c^{\otimes n}(\varrho_m)\big).
\end{eqnarray*}
Combining these last two equations yields
\begin{eqnarray*}
I(A\,\rangle B^nM)_\omega 
&\leq& H(B^n|M)_{\omega'} - H(C^nE^n|M)_{\omega'} \\
&=& H(X^n|M)_{\omega'} - H(C^nE^n|M)_{\omega'} 
\end{eqnarray*}
where we have renamed $B^n$ to $X^n$ to emphasize its classicality.  From now on, we rename $\omega'^{MB^nC^nE^n}$ to $\omega'^{MX^nC^nE^n}$ accordingly.
Identifying $T_i = MX^{i-1}$, $T = ST_S$ and $XCE = X_SC_SE_S$, for $S\sim \text{unif}\{1,\dotsc,n\}$, observe that $S-T-XCE$ forms a Markov chain.  This identification defines the state $\Omega^{TXCE}$, for which 
\begin{eqnarray*}
H(X^n|M)_{\omega'} &=& \sum_{i=1}^n H(X_i|MX^{i-1})_{\om'} \\
&=& n H(X|T)_\Omega.
\end{eqnarray*}
By data processing with respect to appropriate tensor products of the map $\proj{x}\mapsto \psi_x^{CE}$, we may upper bound 
\begin{eqnarray*}
- H(C^nE^n|M)_{\omega'} 
&=& -\sum_{i=1}^n H(E_iC_i|E^{i-1}C^{i-1}M)_{\om'}\\
&\leq& -\sum_{i=1}^nH(C_iE_i|MX^{i-1})_{\om'} \\
&=& - nH(CE|T)_\Omega,
\end{eqnarray*}
obtaining $\frac{1}{n}I(A\,\rangle B^nM)_\omega \leq 
H(X|T)_\Om - H(CE|T)_\Omega$.

It is perhaps instructive to see that $\Omega$ can be explicitly written as
$\Omega^{TXCE} = \bigoplus_t p(t) \Omega^{XCE}_t$,
where we take $\CT = \CM\times\biguplus_s \CX^{s-1}$ (here $\CX^0$ is the empty set), and 
\[\Om^{XCE}_{mx^{s-1}} = \bigoplus_{x_s}p(x_s|m,x^{s-1})\psi_{x_s}^{CE}.\]
We now continue by bounding the mutual information in (\ref{cqconverseI}) via 
\begin{eqnarray*}
I(M;C^n)_\omega &=& I(M;C^n)_{\om'} \\ 
&\leq& nI(T;C)_\Omega.
\end{eqnarray*}
Here, the first step is because $\CN_C^{\otimes n}\circ\deph^{\!\otimes n}= \CN_C^{\otimes n}$, which follows from (\ref{gd1}) because $\CN_C = \tr_E(\CN_B)_c$,
while the second follows from manipulations which are identical to those used to bound
(\ref{nIMC}) in the converse to Theorem~\ref{theo:cqchanneldegraded}; the only differences are that we use data processing with respect to tensor products of the map
$\proj{x}\mapsto \tr_E\psi_x$ and relabel $B^{i-1}$ to $X^{i-1}$.  This proves the claim.
\qed

\section{Quantum region $\CQ(\CN)$ for quantum channels}
\label{section:QQforquantum}
In this next entirely quantum mechanical scenario, Alice attempts to share a large bipartite entangled state with Bob, while also trying to build a large GHZ state with Bob and Charlie.  Alice encodes by preparing the state
$\ket{\Upsilon}^{AGA'^n}$, entangled with the inputs of a large number $n$ of instances of $\CN^{A'\to BC}$. Bob and Charlie employ respective decoding maps $\CD_1^{B^n\to \h{A}G_B}$ and  $\CD_2^{C^n\to G_C}$.  These components comprise a $(Q,Q_B,n,\ep)$ \emph{entanglement generation code} for the broadcast channel $\CN$ if
they generate a rate $Q_B$ EPR state $\ket{\Phi_{Q_B}}^{A\h{A}}$
and a \emph{rate $Q$ GHZ state}
\[\ket{\Gamma_Q}^{GG_BG_C} = \frac{1}{\sqrt{2^{nQ}}}
\sum_{m = 1}^{2^{nQ}}\ket{m}^{G}\ket{m}^{G_B}\ket{m}^{G_C}\]
in the sense that
\[F\Big(\ket{\Phi_{Q_B}}^{A\h{A}}\ket{\Gamma_Q}^{GG_BG_C},
\CN^{\otimes n}(\CD_1\otimes\CD_2)\big(\Upsilon^{GAA'^n}\big)\Big) \geq 1-\ep.\]
Achievable rates and the capacity region $\CQ(\CN)$ are defined in analogy to the earlier scenarios.
In Theorem~\ref{theo:qqisom}, we give a multi-letter formula for $\CQ(\CN)$ in the case where $\CN$ is an isometric channel.  Theorem~\ref{theo:qqdephasing} derives a single-letter formula for $\CQ(\CN)$ in case $\CN$ is an isometric extension of a generalized dephasing channel to Bob.  Note that these results can be regarded as dynamic analogs of those obtained in \cite{svw}, which studies distillation of EPR and GHZ entanglement from arbitrary tripartite pure states.  While those authors allow for additional classical communication, we do not.  Similar correspondences exist in the literature, such as between \cite{dcr} and \cite{dev} for the single-sender/single-receiver case, as well as between \cite{merge} and \cite{qmac} for the case of a single sender and multiple receivers.

\begin{theo}\label{theo:qqisom}
Let $\CU^{A'\rightarrow BC}$ be an isometric broadcast channel.  Then $\CQ(\CU)$ contains the set of pairs of nonnegative quantum rates $(Q,Q_B)$ satisfying 
\begin{eqnarray*}
Q_B &\leq& I(A\,\rangle BT)_\sigma \\
Q &\leq& \min\{I(T;B)_\sigma,I(T;C)_\sigma\} 
\end{eqnarray*}
where $\sigma^{TABC}$ takes the same form as in (\ref{cqarise}), replacing $\CN$ with $\CU$.  The bound on $|\CT|$ is the same as well.  Furthermore, this achievable region regularizes to give the entire capacity region as well.
\end{theo}

\begin{theo} \label{theo:qqdephasing}
Let $\CU^{A'\rightarrow BC}$ be a broadcast channel which is an isometric extension of a generalized dephasing channel to $B$, written
\[\CU = \sum_x \ket{x}^B\ket{\psi_x}^C\bra{x}^{A'}.\]
Then $\CQ(\CU)$ equals the set of pairs of nonnegative quantum rates $(Q,Q_B)$ satisfying 
\begin{eqnarray*}
Q_B &\leq& H(X|T)_\om - H(C|T)_\om \\
Q &\leq& I(T;C)_{\om}
\end{eqnarray*}
where
\[\om^{TXC} = \bigoplus_{x,t} p(t,x) \psi_x^C\]
and $|\CT|\leq |\CX|$.
\end{theo}

When $\CU^{A' \rightarrow BC}$ is an isometric extension of the pinching channel 
$\CP^{A' \rightarrow B}$, this theorem yields the rate region from  Figure~\ref{figure:pinching}
with $R$ replaced by $Q$.  We remark that
by using the standard technique of restricting to a high-fidelity subspace of the input, it is possible to strengthen the previous four theorems to obtain stronger error criteria, such as that from the strong subspace transmission of \cite{qmac}.  We have, however, focused on entanglement generation for simplicity.

\subsection{Quantum coding theorem for quantum channels}
\label{section:QQforquantum:coding}
Here, we will take the codes constructed in Section~\ref{section:CQforquantum:coding} and make the common classical message coherent.  We will use the following two lemmas:
\begin{lem}[Gentle coherent measurement~\cite{hdw05}] \label{lem:gentlecoherent}
Suppose that a POVM $\{\Lambda_m\}$ identifies the elements of a set of pure states $\{\ket{\varphi_m}^{B}\}$, in the sense that $\tr\Lambda_m\varphi_m \geq 1-\ep$ for every $m$.  Then, there is an isometry $\CV^{B\to B\h{B}}$ which satisfies $\bra{m}^{\h{B}}\bra{\varphi_m}\CV\ket{\varphi_m} \geq 
1 - {\epsilon}$
for each $m$.
\end{lem}
\begin{lem} \label{lemma:transitivity}
For any state $\rho^{AB}$ with partial traces $\rho^A$ and $\rho^B$ and any $\ket{\psi}^A$ and $\sigma^B$, we have 
\[F(\rho^{AB},\psi^A\otimes \sig^B) \geq 
1-3\big(1-F(\ket{\psi}^A,\rho^A)\big) - \Norm{\rho^B - \sig^B}_1.\]
\end{lem}
\vspace{.05in}

\noindent {\bf Proof of Theorem~\ref{theo:qqisom}} (Coding theorem).
Letting $\CU^{A'\to BC}$ be an arbitrary isometric broadcast channel, we set $\CN_B = \tr_C\CU$. For any bipartite pure state ensemble $\{p(t),\ket{\phi_t}^{A''A'^n}\}_{m\in 2^{nr}}$ and any $\ep > 0$, the previous coding theorem shows (relabeling $R$ to $Q$ and $Q$ to $Q_B$) that as long as $n$ is large enough, there is a $(Q,Q_B,n,6\sqrt{\ep})$ cq entanglement generation code $\big\{\ket{\Upsilon_m},\CD_m, \{\Lambda_m\},\{\Lambda_m'\}\big\}$ for $\CU^{A'\to BC}$,
provided that the rates satisfy
\[Q < \min\{I(T;B)_\sig,I(T;C)_\sig\}\] and
\[Q_B < I(A''\,\rangle BT)_\sig.\] 
These quantities are computed with respect to the state
\[\sig^{A''BCT} = \bigoplus_t p(t) \CU(\phi_t^{A''A'}).\]
We will show how to make the common classical message coherent.
For each $m\in 2^{nR}$, define
\[\ket{\Upsilon'_m}^{AB^nC^n} = \CU^{\otimes n}\ket{\Upsilon_m}\]
and observe that
\[\bra{\Upsilon'_m}(1^A\otimes\Lambda_m\otimes\Lambda'_m)
\ket{\Upsilon'_m} \geq 1-\ep.\]
By Lemma~\ref{lem:gentlecoherent}, there are thus coherent local measurements
$\CV^{B^n\to B^nG_B}$ and $\CW^{C^n\to C^nG_C}$ satisfying
\begin{equation}
\bra{m}^{G_CG_B}\bra{\Upsilon'_m}(\CV\otimes \CW) \ket{\Upsilon'_m}
\geq 1- \ep\label{eqn:upvwup}
\end{equation}
for each $m$, where we take $\ket{m}^{G_BG_C} \equiv \ket{m}^{G_B}\ket{m}^{G_C}$.
Now, there are local unitaries (permutations of the Hilbert space factors, in fact)
$V_m^{B^n\to B^n}$ and $W_m^{C^n\to C^n}$ which satisfy
\[(V_m\otimes W_m)\ket{\Upsilon'_m} = \ket{\Upsilon'_1}^{AB^nC^n}\]
because $\ket{\Upsilon_m}$ is just a permutation of the $A'^n$ part of
the fixed representative $\ket{\Upsilon_1}^{AA'^n}$.
Define the controlled unitary
\[V^{B^nG_B\to B^nG_B} = \sum_{m} \proj{m}\otimes V_m\]
and similarly define $W^{C^nG_C\to C^nG_C}$.
Setting
\[\ket{\Upsilon''_m}^{AB^nC^nG_BG_C} = \big((V\circ \CV)\otimes (W\circ \CW)\big)\ket{\Upsilon'_m},\]
we may re-express (\ref{eqn:upvwup}) as
\begin{equation}
\bra{m}^{G_BG_C}\bra{\Upsilon'_1}\ket{\Upsilon''_m} \geq 1-\ep.
\label{eqn:mupup}
\end{equation}
We now define Alice's encoding as
\[\ket{\Upsilon}^{GAA'^n} = \frac{1}{\sqrt{2^{nQ}}}\sum_{m} \ket{m}^{G}\ket{\Upsilon_m}^{AA'^n},\]
writing
\[\ket{\Upsilon'}^{GAB^nC^n} = \CU^{\otimes n}\ket{\Upsilon}^{GAA'^n}\]
and
also setting
\[\ket{\Upsilon''}^{GAB^nC^nG_BG_C} = \big((V\circ \CV)\otimes (W\circ \CW)\big)\ket{\Upsilon'}\]
as before. We now bound
\begin{eqnarray}
 \bra{\Gamma_{Q}}^{GG_BG_C}\bra{\Upsilon'_1}^{AB^nC^n}\ket{\Upsilon''}^{GAB^nC^nG_BG_C}
 \hspace{.7in} &  \nn\\
= 2^{-nQ}\sum_{m'm}\bra{m'}^G\bra{m'}^{G_BG_C}\bra{\Upsilon'_1}
\ket{m}^G\ket{\Upsilon''_m} \hspace{.01in} \vphantom{\vrule{\Big|}}\nn\\
= 2^{-nQ}\sum_{m}\bra{m}^{G_BG_C}\bra{\Upsilon'_1}
\ket{\Upsilon''_m}^{AB^nC^nG_BG_C} \nn\\
\geq 1-\ep. \hspace{2.085in} \label{sqrtfid}
\end{eqnarray}
The last line uses the estimate (\ref{eqn:mupup}).
Since the construction in the previous coding theorem guarantees that \begin{equation}
F\big(\ket{\Phi_{Q_B}}^{A\h{A}},\CD_1(\tr_{C^n}\Upsilon'^{AB^nC^n}_1)\big)\geq 1-\ep, \label{F3}
\end{equation}
we may then employ Lemma~\ref{lemma:transitivity} to combine these last two estimates to show that the state 
\[\Omega^{GG_BG_CA\h{A}} = \tr_{C^n}\CD_1(\Upsilon_1''^{GAB^nC^nG_BG_C})\]
which results from the protocol satisfies  
\begin{eqnarray*}
F&\equiv& F\big(\ket{\Gamma_{Q}}^{GG_BG_C}\ket{\Phi_{Q_B}}^{A\h{A}},\Omega^{GG_BG_CA\h{A}}\big) \\
&\geq& 1 - 3\Big(1-F\big(\ket{\Phi_{Q_B}}^{A\h{A}},\Omega^{A\h{A}}\big)\Big) \\
& & - \Norm{\Gamma_{Q}^{GG_BG_C} - \Omega^{GG_BG_C}}_1 \\
 &\geq& 1- 3\ep - \sqrt{8 \epsilon} \\
 &\geq& 1- 6 \sqrt{\ep}. 
\end{eqnarray*}
The bound on the fidelity in the second line is from (\ref{F3}), while the bound on the trace distance in the next line is by application of (\ref{fid2tr}) to the square-root of the fidelity in (\ref{sqrtfid}). This proves the coding theorem. \qed
\subsection{Converse theorems}
Here we complete the proofs of Theorems~\ref{theo:qqisom} and \ref{theo:qqdephasing} by showing that they follow from the converse theorems of Section~\ref{section:CQforquantum:converse}.
\label{section:QQforquantum:converse}

\vspace{.1in}
\noindent {\bf Proofs of Theorems~\ref{theo:qqisom} and \ref{theo:qqdephasing}} (Converses).
Observe that any $(Q,Q_B,n,\ep)$ qq entanglement generation code is able to establish $\ep-$good uniform common randomness between Alice, Bob and Charlie at rate $Q$, in the sense that they generate a triple of random variables $(M_A,M_B,M_C)$ which satisfy 
\[\big| \text{dist}(M_A,M_B,M_C) - \text{dist}(M,M,M)\big|_1 \leq 2\ep,\] 
where $M$ is uniformly distributed on $\{1,\dotsc,2^{nR}\}$.  To accomplish this, Alice will measure the $G$ part of her input $\Upsilon^{GAA'^n}$ in the GHZ basis $\{\ket{m}^G\}$ at any point in the protocol, while Bob and Charlie measure their respective bases
$\{\ket{m}^{G_B}\}$ and $\{\ket{m}^{G_C}\}$ after their decodings are complete.  
The converses for the cq scenario provide upper bounds on this uniform common randomness generation rate for protocols which also generate Alice-Bob entanglement at rate $Q$, therefore proving the converse for Theorem~\ref{theo:qqisom} as well as proving Theorem~\ref{theo:qqdephasing}
\qed

\section{Achievable quantum rates from state merging} \label{section:merging1}
Suppose that three players, Alice, Bob and Charlie, share their parts of many instances of a tripartite pure state $\ket{\psi}^{ABC}$.  
Assuming that Alice can send classical bits to Bob for free, \cite{merge} showed that the quantum communication cost for Alice to transfer her $A^n$ systems to Bob is asymptotically equal to $H(A|B)$, regardless of the negativity of the expression.  Specifically, whenever $H(A|B)$ is negative (or equivalently, when $I(A\,\rangle B) > 0$), Alice and Bob can generate EPR entanglement at rate $I(A\,\rangle B)$ in the process of transferring $A^n$ to Bob, using \emph{only} classical communication and no quantum communication whatsoever.
This is an improvement over standard entanglement distillation \cite{DW05},
where the same amount of EPR entanglement is obtained without deliberately trying to accomplish 
state merging. 
  On the other hand, in case $H(A|B) > 0$, the protocol requires as input Alice-Bob EPR entanglement at a rate of at least  $H(A|B)$ ebits per system to be transferred. Therefore, if Alice and Bob perform an inital state merging with negative cost, they can use the extra entanglement they generate to perform a subsequent merging at an overall positive cost, without investing any entanglement.  As in \cite{merge}, we only directly consider state merging when the cost is negative, as the protocol with positive cost is obtained by having Alice establish an appropriate amount of pure entanglement with Bob, so that the total coherent information they share becomes positive.  Formally, a \emph{negative cost state merging protocol} for a state $\ket{\psi}^{ABC}$ consists of an instrument $\bs{\CM}^{A^n\to DM}$ with components $\CM_m^{A^n\to D}$ to be performed on Alice's systems $A^n$,  together with a collection of decoding operations $\CD_m^{B^n \to B^n \h{A}^n\h{D}}$ for Bob.
The quantum outputs $D$ and $\h{D}$ hold Alice's and Bob's respective halves of the entanglement resulting from the protocol, while
each copy of $\h{A}$ corresponds to a system located in Bob's laboratory which is isomorphic to $A$, whose purpose is to hold the corresponding part of the transferred states.  The protocol proceeds as follows.  Alice performs the instrument $\CM^{A^n\to DM}$ and tells the classical result $M$ to Bob who, depending on that classical data he receives, uses the appropriate decoding map.  These components will be said to comprise a $(Q,n,\ep)$ \emph{negative cost state merging protocol} for $\ket{\psi}^{ABC}$ if $|D| = |\h{D}| = 2^{nQ}$ and
\[F\Big(\ket{\psi}^{\otimes n}\ket{\Phi_Q}^{D\h{D}},
\sum_{m} (1^{C^n}\otimes\CD_m\otimes\CM_m)(\psi^{\otimes n})\Big) \geq 1-\ep,\]  where
$\ket{\Phi_Q}^{D\h{D}}$ is a rate $Q$ maximally entangled state.
The following proposition is from \cite{merge}, and is proved in \cite{merge2}.
\begin{prop} \label{prop:merge}
Let a pure tripartite state $\ket{\psi}^{ABC}$ satisfying $I(A\,\rangle B)_\psi > 0$ be given.  Then, for every $\ep > 0$, and for every $0\leq Q < I(A\,\rangle B)_\psi$, there is $n$ sufficiently large so that there exists a $(Q,n,\ep)$ negative cost state merging protocol for $\ket{\psi}^{ABC}$.
\end{prop}
We now state a theorem.
\begin{theo}
Let $\CN^{A'\to BC}$ be arbitrary.  If Bob can communicate for free with Charlie via a classical channel, then Alice may generate rate $Q_C$ entanglement with Charlie whenever there is a bipartite pure state $\ket{\psi}^{AA'}$
for which
\[Q_C < I(A\,\rangle BC)_\sig \text{ and }
I(B\,\rangle C)_\sig > 0,\]
where $\sigma^{ABC} = \CN(\psi^{AA'})$.
In addition, the same protocol allows Bob and Charlie to generate independent EPR entanglement between themselves at any rate less than $I(B\,\rangle C)_\sigma$.
\end{theo}

\noindent {\bf Proof outline.}
Assume that Alice and Charlie share common randomness.
Fixing a single-letter reference state $\ket{\psi}^{AA'}$ satisfying the conditions of the theorem,
Alice uses a random LSD code of rate $I(A\,\rangle BC)_\sigma$ (see Proposition~\ref{prop:lsd}) based on her
common randomness with Charlie, pretending as though Bob and Charlie can collaborate
in their decoding. As her average code density matrix is close to
the product state $(\psi^{A'})^{\otimes n}$, 
the output state of Bob and Charlie is close to $(\sigma^{BC})^{\otimes n}$.
By assumption, $I(B\,\rangle C)_\sigma > 0$, so there is a \emph{negative cost}
for Bob to transfer his $B^n$ systems to Charlie. This means that Bob and Charlie can distill 
EPR's at any rate less than $I(B\,\rangle C)_\sigma$ during this process. Charlie uses the common randomness
to decode the random LSD code, thus establishing the rate $I(A\,\rangle BC)_\sigma$  entanglement with Alice.
Finally, the protocol is derandomized using standard arguments. \qed

Finally, we demonstrate that Alice may generate, and also \emph{transmit} \cite{bkn} independent entanglement between herself and each
receiver \emph{without} the assistance of classical communication between the two receivers.
\begin{theo}
Let $\CN^{A'\to BC}$ be arbitrary, and let $\ket{\psi}^{A_BA_CA'}$ be entangled between local systems $A_B$ and $A_C$ in Alice's lab and the $A'$ input to the channel. Provided that $I(A_B\,\rangle B)_{\CN(\psi)}$ and $I(A_C\,\rangle C)_{\CN(\psi)}$ are positive, Alice may generate those same amounts of independent entanglement
with each receiver.
\end{theo}

\noindent {\bf Proof outline.}
If communication is allowed between Alice and each of the receivers, the
theorem is immediate from a double application of Proposition~\ref{prop:merge}
(or, rather, entanglement distillation \cite{DW05}, as the state merging aspect is not needed).
We now argue that the classical communication is not needed.
Alice begins such a protocol by applying instruments
$\bs{\CE}^{A_B^n\to \tilde{B}M}$ and ${\bs{\CE}'}^{A_C^n \to \tilde{C}K}$
with components
$\{\CE_m^{A_B^n \rightarrow \tilde{B}}\}_m$ and $\{{\CE'_k}^{A_C^n \to \tilde{C}}\}_k$
to the $A_B$ and $A_C$ parts of the state $\sigma^{A^n_BA^n_CB^nC^n} = (\CN(\psi))^{\otimes n}$.
Conditioned on  receiving the classical message $M = m$, Bob performs
$\CD_m^{B^n \rightarrow \h{B}}$. Conditioned on  receiving the classical message $K = k$, Charlie performs
${\CD'}_k^{C^n \rightarrow \h{C}}$. By entanglement distillation,
there exist $m$ and $k$ such that applying $\CD_m \otimes \CD'_k$ to
$(\CE_m \otimes \CE'_k) (\sigma)/\tr(\CE_m \otimes \CE'_k) (\sigma)$
gives a state close to the tensor product of the two desired maximally entangled states.
Thus, Alice could have prepared
\[\Upsilon^{\hat{B}\hat{C} A'}_{mk} =  (\CE_m \otimes \CE'_k) (\psi^{\otimes n})/\tr(\CE_m \otimes \CE'_k) (\psi^{\otimes n})\]
in the first place, eliminating the need
for classical communication.
If we are interested in entanglement \emph{transmission} instead of
entanglement generation, then Alice is given a purification
of the $\tilde{B}\tilde{C}$ systems rather than  being able to prepare them directly.
Luckily, she may always produce $\Upsilon^{\tilde{B}\tilde{C} A'}_{mk}$ by
a (possibly noisy) encoding $\CF$.
A direct adaptation of the result of \cite{bkn} guarantees that $\CF$ may be replaced by an isometry.
It would be desirable to have a direct proof of this theorem instead of invoking entanglement distillation
and \cite{bkn}.
\qed

We remark that the regularized optimization over such $\ket{\psi}^{A_BA_CA'}$ yields the capacity region when there is no Bob-Charlie communication, although the resulting characterization of this 
capacity region is unlikely to be the most useful.

Rates achievable when there is a positive cost for state merging are closely related to \emph{entanglement-assisted} channel capacities~\cite{BSST}. The structure of entanglement-assisted
capacity regions is often formally similar to their classical counterparts. Indeed, we note that when Alice shares entanglement with both Bob and Charlie, a region formally generalizing Marton's region~\cite{marton} for the broadcast channel becomes achievable~\cite{DHL}.

\section*{Acknowledgements}
JY  is thankful for support from CIFAR through the Department of Computer Science at McGill University, where most of this research was conducted.  He is also thankful for support from the US National Science Foundation under grant PHY-0456720 through the Institute for Quantum Information at the California Institute for Technology, as well as from the LDRD program of the US Department of Energy through Los Alamos National Laboratory, where this manuscript was completed.  
PH is grateful for support from the Canada Research Chairs program, the Perimeter Institute, CIFAR, FQRNT's INTRIQ, NSERC, ONR through grant N000140811249 and QuantumWorks.
 ID was supported by NSF grant no.\ CCF-0524811. 
JY would also like to thank Tom Cover and Young-Han Kim for helpful discussions, as well as Andreas Winter and Aram Harrow for comments on an earlier version of this manuscript.

\section*{Biographies}
Jon Yard received his B.S.\ in Electrical and Computer Engineering from Carnegie Mellon University and his Ph.D.\ in Electrical Engineering in 2005 from Stanford University under the supervision of Tom Cover.  He has since been a Postdoctoral Scholar in Computer Science at McGill University and also in Physics with the Institute for Quantum Information at Caltech.  He has been a postdoc at Los Alamos National Laboratory since 2007 and was awarded a Feynman fellowship there in 2009.  He is especially interested in mathematical aspects of quantum information science.

Patrick Hayden is an associate professor in McGill University's school of computer science and a distinguished research chair of the Perimeter Institute for Theoretical Physics. He is a member of the publications committee of the IEEE Information Theory Society. Prior to joining McGill, Hayden received his doctorate from the University of Oxford in 2001 and was subsequently a postdoctoral fellow at Caltech until 2004. His research focuses on quantum information theory and its applications to other areas of physics and computer science.

Igor Devetak received his Ph.D. in Electrical Engineering from Cornell University in 2002. From 2002 to 2004 he was a post-doctoral researcher at the IBM T.J. Watson Research Center. In 2005 he became an Assistant Professor in the Electrical Engineering department at the University of Southern California. His academic research interests include quantum information theory and quantum error correction. He has been working in the financial sector since 2007.

\appendices
\section{Proof of cardinality bounds for $\CT$}
In Theorem~\ref{theo:supercoding}, let a finite set $\CT$ and conditional probabilities $p(x|t)$ be arbitrary.  Geometrically, this amounts to fixing a $\CT$-labeled set of points on the $\CX$-probability simplex.
We will show that given any probabilities $p(t)$ on $\CT$, there exists another distribution $q(t)$ which puts positive mass on at most $\min\{|\CX|,|B|^2 +|C|^2-1\}$ elements of $\CT$, while satisfying 
\begin{eqnarray*}
I(X;B|T)_q &=& I(X;B|T)_p \\ 
I(T;B)_q &=& I(T;B)_p, \\ 
I(T;C)_q &=& I(T;C)_p, 
\end{eqnarray*}
where the subscript $q$ means the quantity is evaluated on the state \[\rho^{TXBC}_q \equiv \bigoplus_{x,t} q(t)p(x|t)\rho^{BC}_x.\]
By the characterization (\ref{region:qsuper2}) of the superposition coding region, this suffices to give the bound.
We will prove the above with the following lemma:
\begin{lem}[Fenchel and Eggleston \cite{egg}]
Let $\CS\subset\mathbb{R}^n$ have at most $n$ connected components. Then any point in the convexification of $\CS$ can be written as a convex combination of at most $n$ points in $\CS$.  
\end{lem}

It will thus be sufficient to show that the map  
\begin{equation}
f\colon p(t) \mapsto \big(I(X;B|T)_p,I(T;B)_p,I(T;C)_p\big) \nn
\end{equation}
factors through an affine space of sufficiently low dimension.  
To this end, we decompose  
$f$ into a nonlinear part $f_{\text{nl}}$ and an affine part
\begin{eqnarray}
f_{\text{aff}}\colon p(t) &\mapsto& 
\sum_t p(t)
\big(I(X;B|t),\rho^{B}_t,\rho^C_t\big) \nn\\
&=& \big(I(X;B|T),\rho^{B},\rho^C\big), \label{eqn:aff}
\end{eqnarray}
so that the following diagram commutes:
\[\xymatrix{
p(t) \ar[r]^<>(.4){f} \ar[dr]_{f_{\text{aff}}} &
 \big(I(X;B|T)_p,I(T;B)_p,I(T;C)_p\big) \\
& \big(I(X;B|T),\rho^{B},\rho^C\big) \ar[u]_{f_{\text{nl}}}.
}\]
We regard the affine map as producing convex combinations of the points in some affine space parameterizations of the $\big\{\big(I(X;B|t),\rho^{B}_t,\rho^C_t\big)\big\}_{t\in \CT}$, weighted by the probabilities $p(t)$.  As $\rho^B_t$ and $\rho^C_t$ can be specified either by their individual parameterizations or by $p(x|t)$, the more efficient representation requires at most $\min\{|\CX|-1,|B|^2-1 + |C|^2 -1\}$ numbers.  Since the first coordinate can be taken to be $I(X;B|t)$ itself, we see that 
at most $\min\{|\CX|,|B|^2 + |C|^2 - 1\}$ affine parameters are required
to describe (\ref{eqn:aff}).  By continuity, the image of the $\CT$-simplex under $f_{\text{aff}}$ is connected, and so we may use the earlier lemma to infer the existence of probabilities $q(t)$ on $\CT$ with support cardinality at most $\min\{|\CX|,|B|^2 + |C|^2 - 1\}$, while satisfying $f\big(p(t)\big) = f\big(q(t)\big)$.

For Theorem~\ref{theo:cqchanneldegraded}, the degradedness of the channel implies that the $|C|^2-1$ affine parameters of $\rho^C$ depend affinely on those of $\rho^B$, allowing the reduction of the cardinality bound to $|\CT| \leq \min\{|\CX|,|B|^2\}$.

For the bound of Theorem~\ref{theo:cq}, we instead begin by fixing states $\{\rho_t^{A'}\}_{t\in \CT}$.  Here, the affine map outputs convex combinations of the points $(I_c(\rho_t^{A'},\CN_B),\rho_t^B,\rho_t^C)$.
As a parameterization of the possible $\rho_t^B$ and $\rho_t^C$ requires no more than $\min\{|A'|^2-1,|B|^2-1 +|C|^2-1\}$ coordinates, we obtain by similar reasoning as above that it suffices to take $|\CT| \leq \min\{|A'|^2,|B|^2 + |C|^2-1\}$.

The bound for Theorem~\ref{theo:cqdephasing} follows in the same way as the one for Theorem~\ref{theo:cqchanneldegraded}, although the fact that $|B| = |\CX|$ implies that $|\CT|\leq |\CX|$ is sufficient.
In Theorems~\ref{theo:qqisom} and \ref{theo:qqdephasing}, the bounds are the same as those from Theorems~\ref{theo:cq} and \ref{theo:cqdephasing} and follow for the same reasons.

\bibliographystyle{ieeetr}

\end{document}